\documentclass[aps,letterpaper,twocolumn,pra,footinbib,floatfix,showpacs]{revtex4}
\usepackage{amsmath}
\usepackage{amsfonts}
\usepackage{amssymb}
\usepackage[scanall]{psfrag}
\usepackage{psfrag}
\usepackage{graphicx}
\usepackage{color}
\usepackage{subfigure}
\usepackage{bbold}

\usepackage{yypreamble}

\newcommand{\eL}{\mathcal{L}}

\begin{document}

\author{Yariv Yanay}
\affiliation{Laboratory of Atomic and Solid State Physics, Cornell University, Ithaca NY 14850}
\author{Erich J. Mueller}
\affiliation{Laboratory of Atomic and Solid State Physics, Cornell University, Ithaca NY 14850}
\title{Superfluid Density of Weakly Interacting Bosons on a Lattice}
\date{\today}

\begin{abstract}
We use a path integral approach to calculate the superfluid density of a Bose lattice gas in the limit where the number of atoms per site is large. Our analytical expressions agree with numerical results on small systems for low temperatures and relatively weak interactions. We also calculate the superfluid density and drag for two-component lattice bosons. To attain the correct results we develop tools for calculating discrete time path integrals. These tools should be broadly applicable to a range of systems which are naturally described within an overcomplete basis.
\end{abstract}


\maketitle

\section{Introduction}

Superfluidity is one of the most profound collective manifestations of quantum mechanics \cite{Landau1941,Leggett1999}. It is characterized by dissipation-less flow and is analogous to the vanishing resistivity seen in superconductors. The phenomenology of superfluidity is largely contained in Landau's two fluid model: one component, the normal fluid, responds to the motion of the container walls, while the other component, the superfluid, does not. The total density $\rho = \rho_{n} + \rho_{s}$ is the sum of the density of each component. Leggett showed that at zero temperature, in a translationally invariant system, either $\rho_{s}=0$ or $\rho_{n}=0$ \cite{Leggett1998}. In a lattice, however, even at $T=0$, $\rho_{s}/\rho_{n}$ can be finite. Here we calculate the superfluid fraction for an interacting Bose lattice gas in the large filling limit. Our study complements continuum calculations of superfluid densities \cite{Pollock1987,Fil2004,Leggett2001,Dalfovo1999}.

We are largely motivated by experiments of cold bosonic atoms in optical lattices \cite{Greiner2002}. These systems are well described by the Bose-Hubbard model \cite{Fisher1989}, which can be studied using mean field theories \cite{Sheshadri1993} and Quantum Monte Carlo methods \cite{Krauth1991,Wallin1994}. Further motivated by experiments where two bosonic species are trapped on a lattice \cite{Catani2008,Gadway2010,Soltan-Panahi2011}, we also calculate the superfluid density of a two-component system. Such mixtures have rich behavior, including exotic phases such as paired superflow and counter-superflow \cite{Kuklov2003,Kaurov2005}.

To calculate the superfluid fraction we use a functional integral approach where we include quadratic fluctuations about a coherent state which makes the action stationary. This method becomes exact in the weakly-interacting, low-temperature, high-density limit. We give finite temperature results and compare with exact numerical diagonalization on small systems. 

Our calculation involves coherent state path integrals. As was previously established \cite{Wilson2011} there are difficulties with the continuous time limit of these objects. We show explicitly how to calculate the discrete time path integrals. The resulting formalism contains extra terms not seen in the standard approach.

In Section \ref{sec:superfluid} we introduce the physical meaning and thermodynamic definition of the superfluid density. In Section \ref{sec:singlesp} we present the results of the calculation in the case of a single species of bosons on  a lattice, and in Section \ref{sec:2s} we explore the superfluid properties of two-component bosons. Appendix \ref{app:DiscTPT} highlights the necessity of the discrete-time formalism in the use of coherent state path integrals, and Appendix \ref{app:Calcs} demonstrates the technical details of calculating thermodynamic quantities in this formalism.

\section{Superfluid Density}\label{sec:superfluid}

To define the superfluid density $\rho_{s}$ we follow \cite{Leggett1999} and introduce a new thermodynamic variable $\vec v_{s}$ via a thought experiment. We imagine a fluid at rest within an infinitely long cylinder that is itself at rest. This defines the lab frame. We now give the cylinder an infinitesimal velocity $-\vec v_{s}$ along its axis. After we have allowed the container and fluid to reach equilibrium, the mass current as observed in the \emph{cylinder} frame of reference is
\begin{equation} \begin{split}
\vec j = \rho_{s}\vec v_{s},
\end{split} \end{equation}
which defines $\rho_{s}$, the superfluid density. A normal fluid will move as a rigid body with the container and so have $\rho_{s} = 0$; an entirely superfluid liquid will feel no drag and remain at rest in the lab frame, yielding $\rho_{s} = \rho$. It is also convenient to define the normal density,
\begin{equation} \begin{split}
\rho = \rho_{s} + \rho_{n}.
\end{split} \end{equation}

Formally, we may calculate the superfluid density as the second derivative of the free energy density $\mathcal F$ with respect to $\vec v_{s}$,
\begin{equation} \begin{split}
\rho_{s} & = \at{\dee{^{2}\mathcal F}{v_{s}^{2}}}{\vec v_{s} = 0}.
\label{eq:rhosF}
\end{split} \end{equation}
In a more technical language, this indicates that the superfluid density is the low-frequency, long wavelength limit of a transverse current-current correlation function \cite{Baym1969}.

In a translationally invariant system, for a fluid with well-defined quasiparticles, one can express Eq.~(\ref{eq:rhosF}) as a sum over the excitation spectrum, \cite{Pethick2002}
\begin{equation} \begin{split}
\rho_{n} = \intrm{\frac{d^{3}p}{\p{2\pi\hbar}^{3}}}\p{\frac{\vec p \cdot \vec v_{s}}{\abs{\vec v_{s}}}}^{2}\p{-\dee{n_{b}}{\gep_{p}}}_{\vec v_{s}=0}
\label{eq:rhoncont}
\end{split} \end{equation}
where $n_{b} = \br{e^{\gb \E_{k}} - 1}^{-1}$ is the Bose-Einstein distribution function and $\gep_{p}$ is the energy of an excitation of momentum $\vec p$.

In three dimensions, the microscopic understanding of superfluidity involves condensation into a single macroscopically-occupied quantum state. If the wavefunction of that condensed state is given by $\psi\p{\vec r,t} = \sqrt{\rho_{c}\p{\vec r,t}}e^{i\chi\p{\vec r,t}}$, then the superfluid velocity $v_{s}$ is directly related to the phase $\chi$,
\begin{equation} \begin{split}
\vec v_{s} = \frac{\hbar}{m} \grad \chi\p{\vec r,t}.
\end{split} \end{equation}
The variable $\rho_{c}$ defines the condensate fraction, $\rho_{c}/\rho$, the portion of the system that is condensed into the ground state. This fraction is not, in general, equal to the superfluid fraction $\rho_{s}/\rho$.

\subsection*{Experimental probes of $\rho_{s}$}

To measure $\rho_{s}$ in a gas of cold atoms we propose the following experiment. One begins with an equilibrated Bose gas in an optical lattice, confined by an additional harmonic trap. The dimensionality can be controlled by adjusting the intensity of the lattice beams in the relevant directions. The harmonic trap is then turned off, and the lattice accelerated to velocity $v_{s}$ by chirping the frequency of one of the lattice beams. One then turns off the lattice and uses time-of-flight expansion to measure the momentum $p$ of the cloud. In the limit that all steps are adiabatic, the mass contained in the normal component is $p/v_{s}$. Converting this to a density or a superfluid fraction is trivial.

Gadway et al \cite{Gadway2011} have implemented a related protocol, but did not emphasize the fact that they were measuring the superfluid density. Alternate theoretical proposals involve rotation or artificial gauge fields. Ho and Zhou \cite{Ho2009} showed that the superfluid density can be extracted from images of rotating clouds. John, Hadzibabic and Cooper \cite{John2011} identified a global spectroscopic measure of superfluidity, while Carusotto and Castin \cite{Carusotto2011} investigated a local probe.

\section{Single Species}\label{sec:singlesp}

\subsection{Model}

We begin by analyzing the case of a single species of weakly-interacting bosons on an optical lattice. Such as system can be modeled by the single-band Bose-Hubbard Hamiltonian,
\begin{equation} \begin{split}
\hat H & =  -J\sum_{\avg{i,j}} \br{\hat a_{i}\dg \hat a_{j} + \hat a_{j}\dg \hat a_{i}}
\\ &\quad + \sum_{i}\br{ \frac{U}{2} n_{i}\p{\hat n_{i} - 1} - \mu \hat n_{i} }
\label{eq:BH}
\end{split} \end{equation}
where summations are over the sites $i$ and over the pairs of nearest neighbors $\avg{i,j}$. Here $\hat a_{i}$ ($\hat a_{i}\dg$) is the annihilation (creation) operator for a boson on site $i$ and $\hat n_{i}  = \hat a_{i}\dg \hat a_{i}$ is the number operator for the site. In this paper we focus on the case of a cubic $D$-dimensional lattice, taking the lattice spacing to be $a_{0}$ and the volume of the system to be $V$.

The Bose Hubbard model is a good description of the atomic system as long as the band spacing $E_{b}$ is greater than all relevant energy scales in the system, $E_{b}\gg J, U, T$.  Under these conditions, excitation into higher bands can be neglected. In cold atom experiments this spacing scales as $E_{b} \approx \sqrt{4V_{0}E_{R}}$ where  $E_{R}= \frac{\hbar^{2}k^{2}}{2m}$ is the recoil energy for particles of mass $m$ trapped by lasers of wavenumber $k = 2\pi/\gl$, and $V_{0}$ is the optical lattice depth, which is typically of order $V_{0}\sim 10-100 \times E_{R}$. For near-optical lasers and particles lighter than $m\lesssim 100 \unit{amu}$ the single band approximation works up to $T \lesssim 10^{-6} K$ \cite{Jaksch1998}.

We introduce the velocity $v_{s}$ into our model by applying a phase twist $\vec\Delta\Theta$ to the hopping term,
\begin{equation} \begin{split}
\hat a_{i}^{\dg} \hat a_{j} \to e^{-i\vec{\Delta\Theta}\cdot\p{\vec r_{i} - \vec r_{j}}/a_{0}}\hat a_{i}^{\dg} \hat a_{j}
\end{split} \end{equation}
or equivalently, $\hat a_{j} \to e^{i\vec{\Delta\Theta}\cdot\vec r_{j}/a_{0}}\hat a_{j}$, where $\vec r_{i}$ is the position of lattice site $i$. This phase is related to the lattice velocity by
\begin{equation} \begin{split}
\vec v_{s} = \frac{\hbar}{m a_{0}}\vec{\Delta\Theta}
\end{split} \end{equation}
and so we obtain the relation
\begin{equation} \begin{split}
\rho_{s}^{dd\pr} & = \frac{m^{2}a_{0}^{2}}{\hbar^{2}}\br{\dee{^{2}\mathcal F}{\Delta\Theta_{d}\partial\Delta\Theta_{d\pr}}}_{\vec{\Delta\Theta}=0}
\label{eq:rhosdef}
\end{split} \end{equation}
where $d,d\pr = 1,\dotsc, D = x,y,z$ are the lattice directions. In principle, the superfluid density on a lattice may be a symmetric rank 2 tensor, but for the cubic lattice one has $\rho_{s}^{dd\pr} = \gd_{dd\pr}\rho_{s}$.

Like all thermodynamic quantities, the free energy density can be derived from the partition function,
\begin{equation} \begin{split}
\mathcal F = -\frac{1}{V}\frac{1}{\gb}\ln Z,
\label{eq:Fdef}
\end{split} \end{equation}
given by
\begin{equation} \begin{split}
Z = \Tr e^{-\gb \hat H} = \sum_{\ket{\psi}}\bra{\psi}e^{-\gb \hat H}\ket{\psi}
\end{split} \end{equation}
where $\gb = 1/T$ is the inverse temperature and the sum is over a complete set of states $\ket{\psi}$. Introducing the overcomplete coherent state basis, $\hat a_{i}\ket{\rho_{i}, \varphi_{i}} = \sqrt{\rho_{i}}e^{i\varphi_{i}}\ket{\rho_{i}, \varphi_{i}}$, we break up the operator $e^{-\gb \hat H}$ into $N_{t}$ slices and express the partition function as a path integral of the Euclidean action over the classical fields \cite{Altland2010},
\begin{equation} \begin{split}
Z = \oint \mathrm{\mathcal D\rho\mathcal D\varphi}\exp\br{-S_{E}}.
\label{eq:ZSE}
\end{split} \end{equation}

As discussed in Appendix \ref{app:DiscTPT}, one must use the discrete time formulation of the action, 
\begin{equation} \begin{split}
S_{E} = \sum_{t=0}^{N_{t}-1} L_{E}^{t}
\end{split} \end{equation}
with
\begin{equation} \begin{split}
L_{E}^{t} & = \sum_{i}-\log\br{\braket{\rho_{i,t},\varphi_{i,t}}{\rho_{i,t+1},\varphi_{i,t+1}}} 
\\ & \qquad+ \frac{\gb}{N_{t}}\frac{\bra{\rho_{i,t},\varphi_{i,t}} \hat H \ket{\rho_{i,t+1},\varphi_{i,t+1}}}{\braket{\rho_{i,t},\varphi_{i,t}}{\rho_{i,t+1},\varphi_{i,t+1}}}
\\ & = \sum_{i}\frac{\rho_{i,t} + \rho_{i,t+1}}{2} - \sqrt{\rho_{i,t}\rho_{i,t+1}}e^{i\p{\varphi_{i,t+1}-\varphi_{i,t}}}
\\ & \quad - J\Delta t\sum_{\avg{i,j}} \sqrt{\rho_{i,t}\rho_{j,t+1}}e^{i\p{\varphi_{j,t+1}-\varphi_{i,t} - \Delta\Theta_{ji}}}
\\ & \qquad\qquad\qquad + \sqrt{\rho_{j,t}\rho_{i,t+1}}e^{i\p{\varphi_{i,t+1} - \varphi_{j,t} - \Delta\Theta_{ij}}}
\\ &\quad + \frac{U \Delta t}{2}\sum_{i}\rho_{i,\tau}\rho_{i,\tau+1}e^{2i\p{\varphi_{i,\tau+1}-\varphi_{i,\tau}}}
\\ &\quad -  \mu \Delta t\sum_{i}\sqrt{\rho_{i,\tau}\rho_{i,\tau+1}}e^{i\p{\varphi_{i,\tau+1}-\varphi_{i,\tau}}}
\label{eq:LE}
\end{split} \end{equation}
where $\Delta\Theta_{ij} = \vec{\Delta\Theta}\cdot\p{\vec r_{i} - \vec r_{j}}/a_{0}$ and $\Delta t = \gb/N_{t}$ is the discrete time step. We take the number of time steps $N_{t}$ to be large.

\subsection{Saddle-point Approximation}

We expand the fields $\rho_{i}$, $\varphi_{i}$ around the mean density $\bar \rho$ and mean phase twist $\vec{\Delta\Phi} = \sum \Delta\Phi_{d}\hat r_{d}$, with $\hat r_{d}$ the unit vector in direction $d$. For any site $i$ and its nearest neighbors along $d$, $i_{+d}$ and $i_{-d}$, we have
\begin{equation} \begin{split}
\rho_{i,t} & = \bar\rho + \gd\rho_{i,t}
\\ \varphi_{i,t} & = \frac{1}{a_{0}}\vec r_{i}\cdot \vec{\Delta\Phi} + \phi_{i,t}
\\ \varphi_{i_{+d},t} - \varphi_{i,t} & = \Delta\Phi_{d} + \phi_{i_{+d},t} - \phi_{i,t}
\\ \varphi_{i,t} - \varphi_{i_{-d},t} & = \Delta\Phi_{d} + \phi_{i,t} -  \phi_{i_{-d},t}.
\end{split} \end{equation}

We take these perturbations to be small, $\gd\rho_{i,t}\ll \bar\rho$, $\phi_{i_{\pm d},t} - \phi_{i,t}\ll 1$, $\phi_{i,t+1} - \phi_{i,t}\ll 1$. The validity of these assumptions is examined below, in Sec. \ref{sec:valid}. In particular, when $T,U\lesssim \bar \rho J$ one finds $\avg{\gd\rho_{i,t}^{2}}\sim\bar\rho$ and $\avg{\p{\phi_{x+1}-\phi_{x}}^{2}}\lesssim 1/\bar\rho$. Thus if $\bar\rho \ll 1$ this expansion is well behaved.

Although we assume $\p{\phi_{i_{\pm d},t} - \phi_{i,t}}$ and $\p{\phi_{i,t+1} - \phi_{i,t}}$ are small, we make no assumption that $\phi_{i,t}$ itself is small. Consequently our calculation is valid even in low dimensions, where the condensate fraction vanishes and there is no long range order.

Eq.~(\ref{eq:LE}) expanded around the mean values reads
\begin{equation} \begin{split}
L_{E}^{t} & = \sum_{i}\mathcal L_{0} + \mathcal L_{1}^{i,t} + \mathcal L_{2}^{i,t} + \mathcal L_{int}^{i,t},
\end{split} \end{equation}
where each subsequent term involves higher powers of the fluctuations.

The first term is a constant,
\begin{equation} \begin{split}
\eL_{0} & = \bmat{-\sum_{d} 2 \bar\rho J \cos\p{\Delta\Phi_{d} - \Delta\Theta_{d}} 
\\ + \frac{U}{2}\bar\rho^{2} - \mu \bar\rho}\Delta t.
\label{eq:L0}
\end{split} \end{equation}
Keeping only this term gives the mean-field Gross-Pitaevskii approximation where $\rho_{s} = \bar\rho = \rho$.

The second term, linear in the perturbation, is
\begin{equation} \begin{split}
\eL_{1}^{i,t} = & \bmat{-2J \sum_{d}\cos\p{\Delta\Phi_{d} - \Delta\Theta_{d}} \\ 
+ U\bar\rho - \mu}\Delta t \gd\rho_{i,t}.
\end{split} \end{equation}

The saddle-point mean values minimizing $\eL_{0}$ are
\begin{equation} \begin{split}
& \Delta\Phi = \Delta\Theta
\\ \bar\rho = \frac{1}{U}&\p{\mu + 2J\sum_{d}\cos\p{\Delta\Phi_{d}}}.
\end{split} \end{equation}
Setting $\bar\rho$ to this value makes $\eL_{1}$ vanish. Such a structure is generic, as minimizing the zeroth-order action causes the first order action to vanish. To calculate the superfluid density, we take $\vec{\Delta\Theta} = 0$ but keep $\vec{\Delta\Phi}$ finite, giving the bosons velocity $\frac{\hbar}{ma_{0}}\vec{\Delta\Phi}$ relative to the lattice. The superfluid density becomes $\rho_{s} = \frac{m^{2}a_{0}^{2}}{\hbar^{2}}\br{\dee{^{2}\mathcal F}{\Delta\Phi_{d}^{2}}}_{\vec{\Delta\Phi} = 0}$.

The ``interaction'' term, which we neglect in our calculations, consists of terms of third order or higher in the perturbation fields,
\begin{equation} \begin{split}
\eL_{int}^{i}/\bar\rho = O\p{\gd\rho/\bar\rho,\phi_{j}-\phi_{i}}^{3} = O\p{1/\sqrt{\bar\rho}}^{3}.
\label{eq:Lint}
\end{split} \end{equation}

Our non-trivial results come from the the quadratic term, which is best expressed in momentum space,
\begin{equation} \begin{split}
S_{E} = &\; \sum_{n}  \intrm{\frac{a_{0}^{D}d^{D}k}{\p{2\pi}^{D}}} \br{\frac{V}{{a_{0}}^{D}}\eL_{0} + \eL_{2}^{\vk,\omega_{n}} + \eL_{int}^{\vk,\omega_{n}}}
\label{eq:SEk}
\end{split} \end{equation}
where summation is over $n=-\frac{N_{t}-1}{2}\dotsc\frac{N_{t}-1}{2}$ with frequencies given by $\omega_{n} = \frac{2\pi}{\gb}n$, and the integration is over the first Brillouin zone $\abs{k_{d}}\le \pi/a_{0}$.

Appendix \ref{app:Calcs} provides details on the explicit form of $\eL_{2}^{\vk,\omega}$ and the calculation of propagators. However, all significant physical results rely only on the behavior of the propagators and action at two regimes: $\omega\Delta t\ll 1$ (superscript $p$ for pole behavior) and $\omega\Delta t = \pi e^{i\chi}$ (superscript $\circ$ for contour behavior). These are given, at $\Delta\Phi = 0$, by
\begin{widetext}
\begin{equation} \begin{split}
\avg{\gd\rho\gd\rho}^{p}_{\vec k,\omega} = \frac{V}{a_{0}^{D}} \bar\rho\br{\frac{1}{\Delta t}\frac{2 \E_{1k}}{\omega^{2} + {\E_{k}}^{2}} + O\p{\Delta t}^{0}}
& \quad\quad \avg{\gd\rho\gd\rho}^{\circ}_{\vec k,\omega}  = \frac{V}{a_{0}^{D}} \bar\rho\br{1 + \frac{\E_{1k}\Delta t}{1 - \cos\p{\pi e^{i\chi}}}  + O\p{\Delta t}^{2}}
\\ \avg{\gd\rho\phi}^{p}_{\vec k,\omega}  = \frac{V}{a_{0}^{D}}\br{-\frac{1}{\Delta t}\frac{\omega}{\omega^{2} + {\E_{k}}^{2}} + O\p{\Delta t}^{0}}
& \quad\quad \avg{\gd\rho\phi}^{\circ}_{\vec k,\omega}  = \frac{V}{a_{0}^{D}}  \br{-\half\frac{\sin\p{\pi e^{i\chi}}}{1 - \cos\p{\pi e^{i\chi}}} + O\p{\Delta t}^{2}}
 \\ \avg{\phi\phi}^{p}_{\vec k,\omega}   = \frac{V}{a_{0}^{D}} \frac{1}{4\bar\rho}\br{\frac{1}{\Delta t}\frac{2 \E_{2k}}{\omega^{2} + {\E_{k}}^{2}} + O\p{\Delta t}^{0}}
& \quad\quad \avg{\phi\phi}^{\circ}_{\vec k,\omega}  = \frac{V}{a_{0}^{D}} \frac{1}{4\bar\rho}\br{1 + \frac{\E_{2k}\Delta t}{1 - \cos\p{\pi e^{i\chi}}} + O\p{\Delta t}^{2}},
\label{eq:proppole}
\end{split} \end{equation}
\end{widetext}
where we use the notation $\avg{XY}_{\vk,\omega} = \avg{X_{\vk,\omega} Y_{-\vk,-\omega}}$ and on the right it is understood $\omega = \frac{\pi}{\Delta t}e^{i\chi}$. The energies appearing in these expressions are
\begin{equation} \begin{split}
\E_{1k} & = 4J\sum_{d}\sin^{2}\p{k_{d}a_{0}/2}, \\ \E_{2k} &= 2 \bar\rho U + \p{4J\sum_{d}\sin^{2}\p{k_{d}a_{0}/2}}
\\  {\E_{k}}^{2} & = \br{4J\sum_{d}\sin^{2}\p{k_{d}a_{0}/2}} \times
\\& \quad \quad \br{2 \bar\rho U + \p{4J\sum_{d}\sin^{2}\p{k_{d}a_{0}/2}}}
\label{eq:Ek}
\end{split} \end{equation} 
In the continuum limit, one has $\E_{1k} \to \frac{\hbar^{2}\vk^{2}}{2m}$ and $\E_{2k} \to \frac{\hbar^{2}\vk^{2}}{2m} + 2 g \rho$, where $g = U a_{0}^{D}/m$, $\rho = m \avg{n}/a_{0}^{D}$. The excitation spectrum $\E_{k}$ then corresponds to the familiar Bogoliubov result.

\subsection{Superfluid Density}
The superfluid density is given by
\begin{equation} \begin{split}
\rho_{s} = \frac{m^{2}a_{0}^{2}}{\hbar^{2}}\p{-\frac{1}{\gb V}}\br{\dee{^{2}\ln Z}{\Delta\Phi_{d}^{2}}}_{\vec{\Delta\Phi} = 0}.
\end{split} \end{equation}
Some insight may be gained by inserting the path integral expressions for the free energy density and the partition function into this equation. We find that the superfluid density, to order $O\p{1/\bar\rho}^{0}$, can be decomposed into three terms,
\begin{equation} \begin{split}
\rho_{s} = \frac{2ma_{0}^{2}J}{\hbar^{2}} \frac{m}{a_{0}^{D}}\br{n_{0} - n_{n}^{U} - n_{n}^{\rho\phi}}.
\label{eq:rhos}
\end{split} \end{equation}
We identify the first term as the total density, from which two normal-density terms are subtracted.

These are, respectively, 
\begin{equation} \begin{split}
\frac{2 J}{a_{0}^{D}} &n_{0}  = -\frac{1}{\gb V}\frac{N_{t} V}{a_{0}^{D}}\avg{\at{\dee{^{2}\eL_{0}}{\Delta\Phi_{d}^{2}}}{\bar\rho}}
\\  & - \frac{1}{\gb V}\sum_{n}\intrm{\frac{a_{0}^{D}d^{D}k}{\p{2\pi}^{D}}} \dee{^{2}\bar\rho}{\Delta\Phi_{d}^{2}}\avg{\at{\dee{\eL_{2}^{\vk,\omega_{n}}}{\bar\rho}}{\Delta\Phi}},
\end{split} \end{equation}
\begin{equation} \begin{split}
\frac{2 J}{a_{0}^{D}}n_{n}^{U} = \frac{1}{\gb V}\sum_{n}\intrm{\frac{a_{0}^{D}d^{D}k}{\p{2\pi}^{D}}} \avg{\at{\dee{^{2}\eL_{2}^{\vk,\omega_{n}}}{\Delta\Phi_{d}^{2}}}{\bar\rho}},
\end{split} \end{equation}
\begin{equation} \begin{split}
&\frac{2 J}{a_{0}^{D}} n_{n}^{\rho\phi}  =  \frac{1}{\gb V}\sum_{m,n}\intrm{\frac{a_{0}^{D}d^{D}k}{\p{2\pi}^{D}}\frac{a_{0}^{D}d^{D}q}{\p{2\pi}^{D}}}
\\  &\footnotesize{\avg{\at{\dee{\eL_{2}^{\vk,\omega_{n}}}{\Delta\Phi_{d}}}{\bar\rho}\at{\dee{\eL_{2}^{\vec q,\omega_{m}}}{\Delta\Phi_{d}}}{\bar\rho}} 
- \avg{\at{\dee{\eL_{2}^{\vk,\omega_{n}}}{\Delta\Phi_{d}}}{\bar\rho}}\avg{\at{\dee{\eL_{2}^{\vphantom{k}\vec q,\omega_{m}}}{\Delta\Phi_{d}}}{\bar\rho}}}.
\end{split} \end{equation}
where $\avg{X} = \frac{1}{Z}\oint \mathrm{\mathcal D\rho\mathcal D\varphi} X\exp\br{S_{E}}$; on the right-hand side of the equations, derivatives in $\Delta\Phi_{d}$  and $\bar\rho$ are to be taken at a constant $\bar\rho$  and $\vec{\Delta\Phi}$, respectively, and then evaluated at $\vec{\Delta\Phi}=0$; and we have omitted multiple vanishing terms. This is similar to the calculation shown explicitly in Appendix \ref{app:Calcs}.

The term $n_{0} = \avg{n} = -\dee{\mathcal F}{\mu}$ is the total average occupation number. It is given by
\begin{equation} \begin{split}
n_{0} & = \bar\rho + \half\intrm{\frac{{a_{0}}^{D}d^{D}k}{\p{2\pi}^{D}}}\p{1 - \frac{\E_{1k}}{\E_{k}}\coth\p{\gb \E_{k}/2}}.
\label{eq:n0}
\end{split} \end{equation}
At $T = 0$, one finds $n_{0} \to \p{\mu + 2D J}/U$ as $U/J\to 0$, and $n_{0}\to \mu/U + \half$ as $U/J\to\infty$. These correspond to the correct occupation numbers in the non-interacting and the no-hopping regimes. The explicit calculation of this term is given in Appendix \ref{app:Calcs}.

The term $n_{n}^{\rho\phi}$ is given by
\begin{equation} \begin{split}
n_{n}^{\gd\rho\phi} & = \intrm{\frac{{a_{0}}^{D}d^{D}k}{\p{2\pi}^{D}}} 2J \sin^{2}\p{k_{d}a_{0}}\p{-\dee{n_{b}}{\E_{k}}}_{\Delta\Phi = 0}.
\label{eq:nRF}
\end{split} \end{equation}
This expression is reminiscent of the form of the normal density in the continuum case, given in Eq.~(\ref{eq:rhoncont}), and it likewise vanishes at $T = 0$.

The additional term, $n_{n}^{U}$, can be understood to come from density-density and phase-phase correlations created by the interaction term in the hamiltonian.
\begin{equation} \begin{split}
n_{n}^{U}  &= \intrm{\frac{{a_{0}}^{D}d^{D}k}{\p{2\pi}^{D}}} \p{1 - \cos\p{k_{d}a_{0}}} \times
\\ & \qquad\qquad\quad  \half \br{\frac{\p{\E_{1k} + \E_{2k}}}{2\E_{k}}\coth\p{\gb \E_{k}/2} - 1}.
\label{eq:nU}
\end{split} \end{equation}
At $T = 0$, $U=0$, this term vanishes and the superfluid fraction becomes one; at non-zero values of $U$ this term is finite even at $T = 0$.

\begin{figure}[htbp] 
   \centering
   \includegraphics[width=\columnwidth]{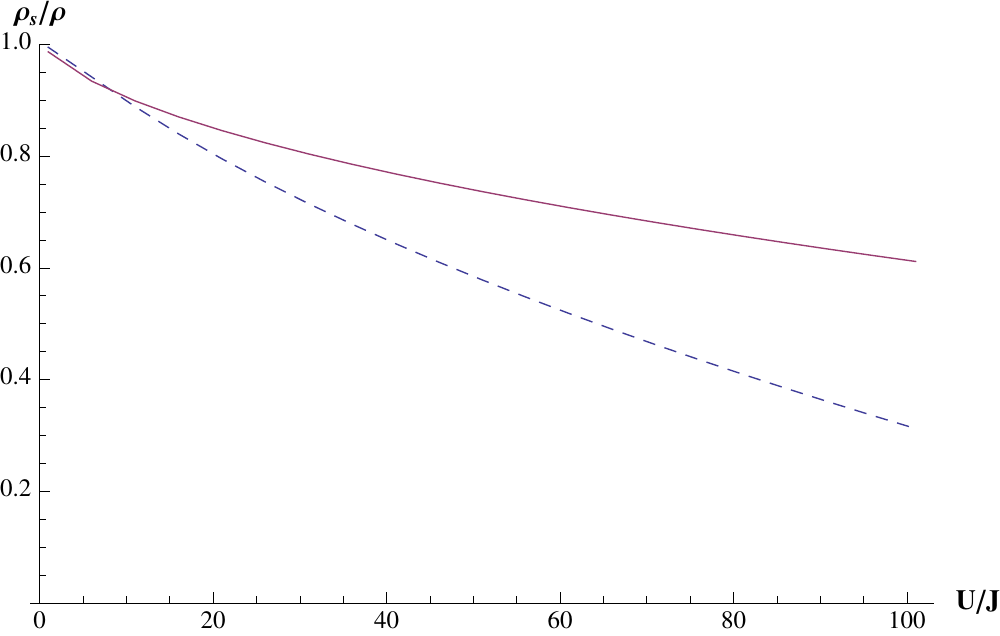} 
   \caption{(Color online) The superfluid fraction $\rho_{s}/\rho$ as function of $U/J$ in an infinite 3D cubic lattice, for $\avg{n} = 10$ (solid red line) and $\avg{n}=1$ (dashed blue line), calculated to leading order in a $1/\avg{n}$ expansion. As discussed in the text, the results are not expected to be quantitatively accurate above $U/J \gtrsim \avg{n}$.}
   \label{fig:rhos3DUJ}
\end{figure}

\begin{figure}[htbp] 
   \centering
   \includegraphics[width=\columnwidth]{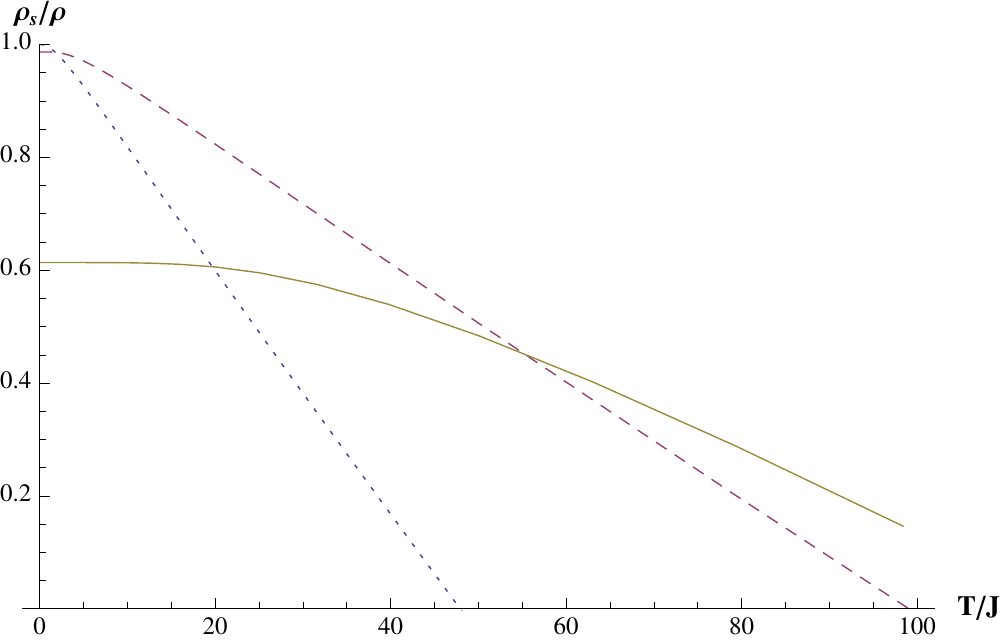} 
   \caption{(Color online) The superfluid fraction $\rho_{s}/\rho$ as function of $T/J$ in an infinite 3D cubic lattice, for $\avg{n} = 10$, at $U/J=0.01$ (dotted blue line), $U/J=1$ (dashed red line) and $U/J=100$ (solid yellow line). At $U=0$, $\rho_{s}$ vanishes at $T/J = 41.5$, the ideal gas transition temperature. As discussed in the text, the results are not expected to be quantitatively accurate at $T/\sqrt{J\p{J + \bar\rho U}} \gtrsim \avg{n}$.}
   \label{fig:rhos3DT}
\end{figure}

\begin{figure}[htbp] 
   \centering
   \includegraphics[width=\columnwidth]{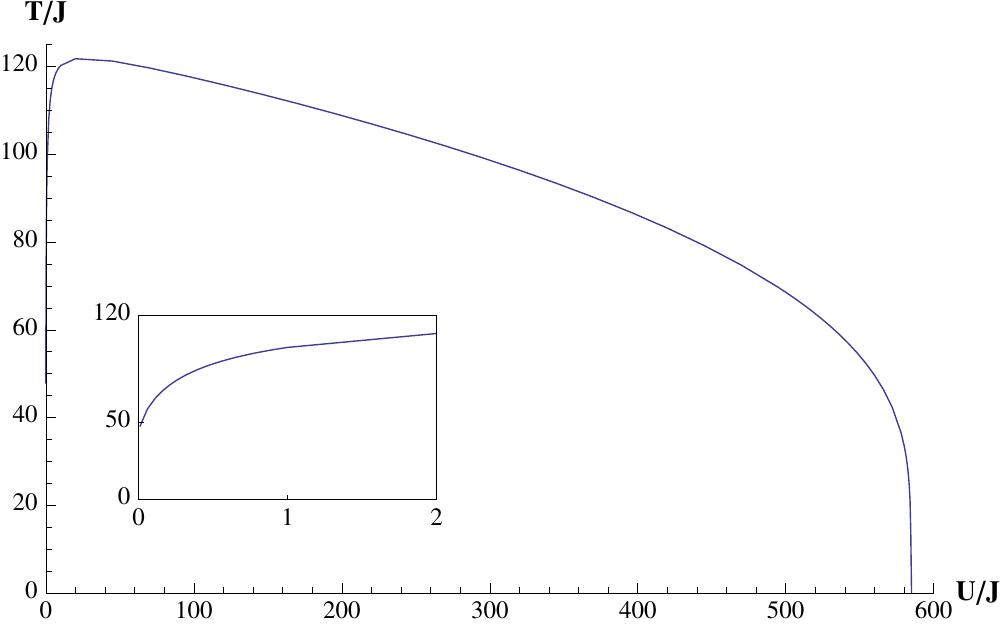} 
   \caption{The values of $U/J, T/J$ at the intercept $\rho_{s} = 0$, suggesting a superfluid-Mott insulator transition. The calculation is performed for an infinite 3D cubic lattice, with $\avg{n} = 10$. The inset shows the form of the curve at small $T/J$.
   As discussed in the text, the results are not expected to be quantitatively accurate at values of $U/J \gtrsim \avg{n}$ or $T/\sqrt{J\p{J + \bar\rho U}} \gtrsim \avg{n}$, but the form is qualitatively similar to curves generated by quantum Monte Carlo methods \cite{Capogrosso-Sansone2007}. }
   \label{fig:rhoseq0TU}
\end{figure}

The resulting superfluid fraction, $\rho_{s}/\rho$, is plotted in Fig.~\ref{fig:rhos3DUJ} at zero temperature as a function of $U/J$ and in Fig.~\ref{fig:rhos3DT} for set values of $U/J$ as a function of temperature. In Fig.~\ref{fig:rhoseq0TU} we show the curve $U\p{T}$ where $\rho_{s}$ vanishes, suggesting a phase transition. The limits of validity of these results will be discussed in Sec.~\ref{sec:valid}.


\subsection{Analytical Limits}
Here we examine the behavior of Eq.~(\ref{eq:rhos}) in several limiting cases. 

First we compare our result to the continuum limit by taking $a_{0}\to 0, J\to\infty$ so that $J a_{0}^{2}$ is constant. In this case, the second term Eq.~(\ref{eq:nU}) vanishes. This can be seen by separately considering the contributions from $k\sim 1/a_{0}$ and $k\ll 1/a_{0}$; in both cases the integrand vanishes. The first part of the normal density, Eq.~(\ref{eq:nRF}), has contribution only from finite momenta, and becomes
\begin{equation} \begin{split}
\frac{m}{a_{0}^{D}}n_{n}^{\gd\rho\phi} & \to 2mJa_{0}^{2}\intrm{\frac{d^{D}k}{\p{2\pi}^{D}}}  k_{d}^{2}\p{-\dee{n_{b}}{\E^{c}_{k}}}_{\Delta\Phi = 0},
\end{split} \end{equation}
with $\E^{c}_{k}$ the continuum spectrum. Identifying $2m J a_{0}^{2}/\hbar^{2} \to 1$, this is precisely the known continuum result seen in Eq.~(\ref{eq:rhoncont}).

Another important limit is zero temperature and large $\bar\rho$. If $U\sim J$, then $\bar\rho U \gg J$ and to first order $\E_{k} \approx \sqrt{8U J \bar\rho\p{\sum_{d}\sin^{2}\p{k_{d}a_{0}/2}}}$, $\E_{1k} + \E_{2k} \approx 2\bar\rho U$ and the normal density becomes
\begin{equation} \begin{split}
n_{n}^{U}  \approx &\;\intrm{\frac{{a_{0}}^{D}d^{D}k}{\p{2\pi}^{D}}} \p{1 - \cos\p{k_{d}a_{0}}}  \half \br{\frac{\p{\bar\rho U}}{\E_{k}} - 1}
\\ & \approx \bar\rho\br{\sqrt{\frac{1}{f_{D}}\frac{U}{J\bar\rho}}} - \half
\end{split} \end{equation}
where $\frac{1}{\sqrt{f_{D}}} = \intrm{\frac{d^{D}\theta}{\p{2\pi}^{D}}} \frac{\p{1 - \cos\p{\theta_{x}}}}{\sqrt{32\p{\sum_{d}\sin^{2}\p{\theta_{d}/2}}}} \approx \frac{1}{\sqrt{20}},\frac{1}{\sqrt{35}},\frac{1}{\sqrt{51}}$ in one, two and three dimensions respectively. This expression is suggestive of a phase  transition from superfluid to Mott insulator at $U=U_{c} \sim f_{D} \bar\rho J$. In two and three dimensions, the values for $f_{D}$ are about double the mean-field result of $U_{c} \sim 2 D\times 4 \p{\bar n + \half}$ \cite{VanOosten2001}. More comparisons along these lines are made in Sec. \ref{sec:Gutz}. As discussed in Section \ref{sec:valid}, these estimates are beyond the range of $U/J$ where our approximations are quantitatively valid. It is nonetheless appealing to see the Mott transition appearing within this formalism.

Finally we consider the free particle case $U=0$. There we have as a function of $T$
\begin{equation} \begin{split}
&n_{n}  = n_{n}^{U} + n_{n}^{\rho\phi} = 
\\ & \quad  \intrm{\frac{{a_{0}}^{D}d^{D}k}{\p{2\pi}^{D}}} \half\p{\coth\p{\gb \E_{k}/2} - 1}
\\ &  + \intrm{\frac{{a_{0}}^{D}d^{D}k}{\p{2\pi}^{D}}} 
\half \bmat{\cos\p{k_{d}a_{0}} \p{1 - \coth\p{\gb \E_{k}/2}} \\+ \frac{\gb J \sin^{2}\p{k_{d}a_{0}}}{\sinh^{2}\p{\gb\E_{k}/2}}}.
\end{split} \end{equation}
The integrand in the first line $\half\p{\coth\p{\gb \E_{k}/2} - 1} = n_{b}\p{\E_{k}}$ and the one in the second is a total derivative that vanishes at $k_{d}a_{0} = 0,2\pi$, and so  $n_{n} = n_{ex}$, the total occupation of excited states. $\rho_{s}$ vanishes at $\avg{n} = n_{ex}$, corresponding to the ideal gas transition temperature.

\subsection{Realm of Validity}\label{sec:valid}

Though the formulation of the action in Eq.~(\ref{eq:LE}), (\ref{eq:SEk}) is exact, our calculations are performed by neglecting the infinite series of terms in $\eL_{int}^{\vk,\omega}$. We can place bounds on the realms of validity of this approximation by requiring that the perturbations from the mean values $\bar\rho$, $\Delta \Phi$ be small,
\begin{equation} \begin{split}
\avg{\gd\rho_{i,t}\gd\rho_{i,t}}  \lesssim & \;\bar\rho^{2}
\\ \avg{\p{\phi_{i_{+d},t} - \phi_{i,t}}^{2}} & \lesssim 1
\\ \avg{\p{\phi_{i,t+1} - \phi_{i,t}}^{2}} & \lesssim 1.
\end{split} \end{equation}
We do not require the phases themselves to be small, only the deviation from one site to another and from one time step to another.

These fluctuations can be calculated by the use of the propagators in Eq.~(\ref{eq:proppole}),
\begin{equation} \begin{split}
\avg{\p{\phi_{i,t+1} - \phi_{i,t}}^{2}} = \frac{1}{2\bar\rho},
\end{split} \end{equation}
\begin{equation} \begin{split}
& \avg{\p{\phi_{i_{+d},t} - \phi_{i,t}}^{2}} =
\\ &\quad\quad  \frac{1}{4\bar\rho}\br{2 + \intrm{\frac{a_{0}^{D}d^{D}k}{\p{2\pi}^{D}}} \frac{\E_{k}}{J} \coth\p{\gb\E_{k}/2}},
\label{eq:dphidphi}
\end{split} \end{equation}
\begin{equation} \begin{split}
\frac{\avg{\gd\rho_{i}\gd\rho_{i}}}{\bar\rho^{2}} & = \frac{1}{\bar\rho}\br{1 + \intrm{\frac{a_{0}^{D}d^{D}k}{\p{2\pi}^{D}}} \frac{\E_{1k}}{\E_{k}} \coth\p{\gb\E_{k}/2}}.
\label{eq:drhodrho}
\end{split} \end{equation}
Examination of the integrals in the latter two inequalities implies that to to keep these parameters small we must have
\begin{equation} \begin{split}
&\quad \bar\rho \gtrsim 1
\\ &U/J  \lesssim \bar\rho
\\ &T/J  \lesssim \bar\rho
\\ T/&\sqrt{J\p{J + \bar\rho U}} \lesssim \bar\rho,
\label{eq:valid}
\end{split} \end{equation}
where the first constraint is universally required, the second stems from the density fluctuations in Eq.~(\ref{eq:drhodrho}) and the last two from the phase fluctuations in Eq.~(\ref{eq:dphidphi}).

\subsection{Gutzwiller Ansatz}\label{sec:Gutz}

At zero temperature, an alternative approach to calculating the superfluid density is to use the Gutzwiller ansatz \cite{Krauth1992}. This is an uncontrolled variational method which reproduces the Bogoliubov results at weak coupling \cite{Sheshadri1993}, and gives us a point of comparison for our results.

The Gutzwiller approach assumes that the ground state of the lattice system may be decomposed into a product of single-site states,
\begin{equation} \begin{split}
\ket{\psi_{G}} = \prod_{i}\ket{g}_{i}
\end{split} \end{equation}
where
\begin{equation} \begin{split}
\ket{g}_{i} = \sum_{m=0}^{\infty}\ga_{i}^{m}e^{i\theta^{m}_{i}}\p{a_{i}\dg}^{m}\ket{0}_{i}
\end{split} \end{equation}
and $\ket{0}_{i}$ denotes the zero-boson state of site $i$. One finds the coefficients $\ga^{m}_{i}, \theta^{m}_{i}$ by minimizing the expectation value $\bra{\psi_{G}}\hat H\ket{\psi_{G}}$. We expect $\ga^{m}_{i} = \ga^{m}$ to be equal on all sites and $\theta^{m}_{i} = \p{\vec{\Delta\Phi} \cdot \vec r_{i}/a_{0}} m$ where $\Delta\Phi$ is the phase twist, as before.

One then minimizes
\begin{equation} \begin{split}
\bra{\psi_{G}}\hat H&\ket{\psi_{G}}  = 
\\ \sum_{i}& - 2J  \sum_{d}\cos\p{\Delta \Phi_{d}}\times
\\ &\qquad\; \p{\sum_{m}\sqrt{m+1}\ga^{m+1}\ga^{m}}^{2}
\\&+ \sum_{m}\p{\frac{U}{2}m\p{m-1} - \mu m}\abs{\ga^{m}}^{2}
\label{eq:GutzE}
\end{split} \end{equation}
and finds that the superfluid density is
\begin{equation} \begin{split}
\rho_{s} & = \frac{m^{2}a_{0}^{2}}{\hbar^{2}}\frac{1}{V}\br{\dee{^{2} }{\Delta\Phi_{d}^{2}} \bra{\psi_{G}}\hat H\ket{\psi_{G}}}_{\vec{\Delta\Phi} = 0}
\\ & = \frac{2m a_{0}^{2}J}{\hbar^{2}}\frac{m}{a_{0}^{D}}\p{\sum_{m}\sqrt{m+1}\ga^{m+1}\ga^{m}}^{2}.
\end{split} \end{equation}

\begin{figure}[htbp] 
   \centering
   \includegraphics[width=\columnwidth]{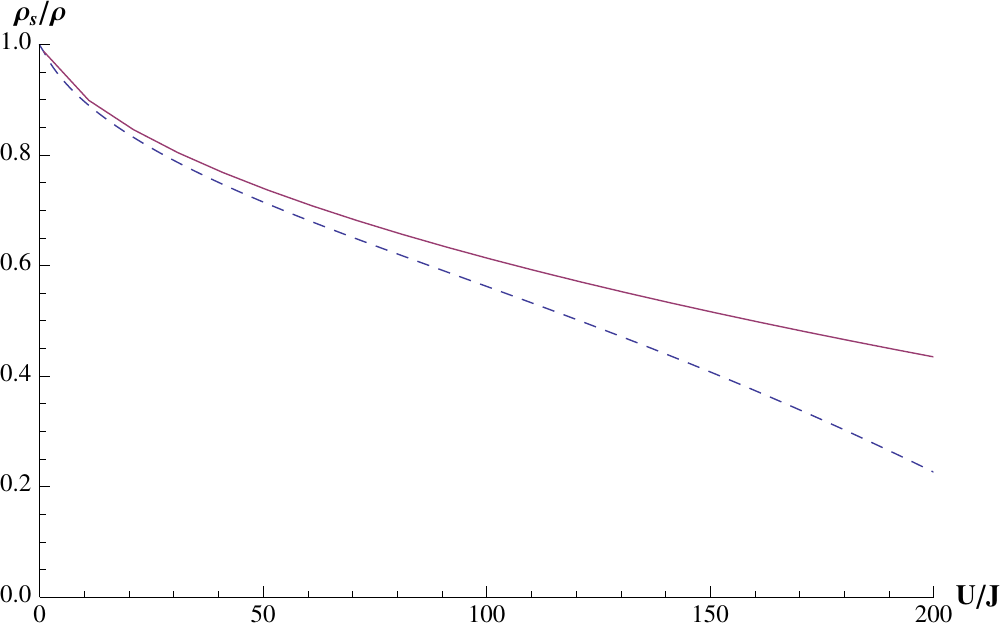} 
   \caption{(Color online) The superfluid fraction $\rho_{s}/\rho$ for an infinite three-dimensional cubic lattice with $\avg{n} = 10$, at $T=0$. The dashed blue line shows the result of the Gutziller-ansatz calculation and the solid red line shows result as calculated using Eqs.~(\ref{eq:rhos})-(\ref{eq:nU}).}
   \label{fig:Gutziller}
\end{figure}

We calculated the parameters $\ga^{m}$ numerically by cutting off the sum at $m=20$. We compare the results with those of Eqs.~(\ref{eq:rhos})-(\ref{eq:nU}) in Fig.~\ref{fig:Gutziller}.

\subsection{Numerical Comparison}

We also compared the results of Eqs.~(\ref{eq:rhos})-(\ref{eq:nU}) to an exact numerical calculation of the superfluid density for a variety of small lattices in one and two dimensions. For a finite lattice and fixed number of particles, we can represent the Hamiltonian in Eq.~(\ref{eq:BH}) as a finite matrix. We diagonalized this matrix, finding all
 eigenstates and eigenvalues. We calculated the superfluid density by performing the full weighted trace over all eigenstates.

\begin{figure}[htbp] 
   \centering
   \includegraphics[width=\columnwidth]{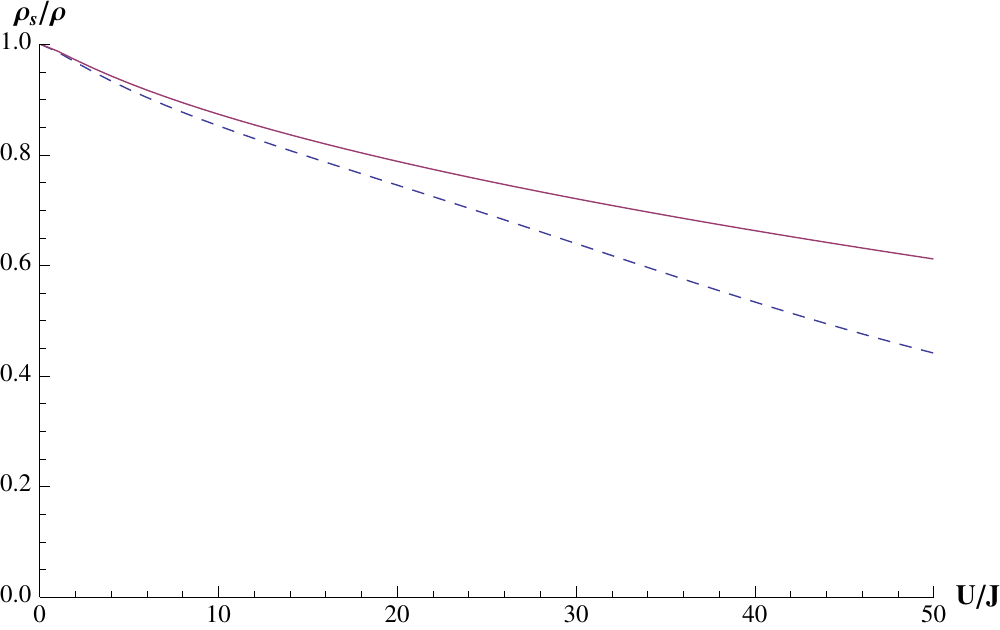} 
   \caption{(Color online) The superfluid fraction $\rho_{s}/\rho$ for a two-dimensional two-by-two lattice with 16 particles, at $T=0$. The dashed blue line shows the numerically exact result and the solid red line shows the analytic approximation.}
   \label{fig:2DT0}
\end{figure}

\begin{figure}[htbp] 
   \centering
   \subfigure[$U/J = 1$]{
   \includegraphics[width=\columnwidth]{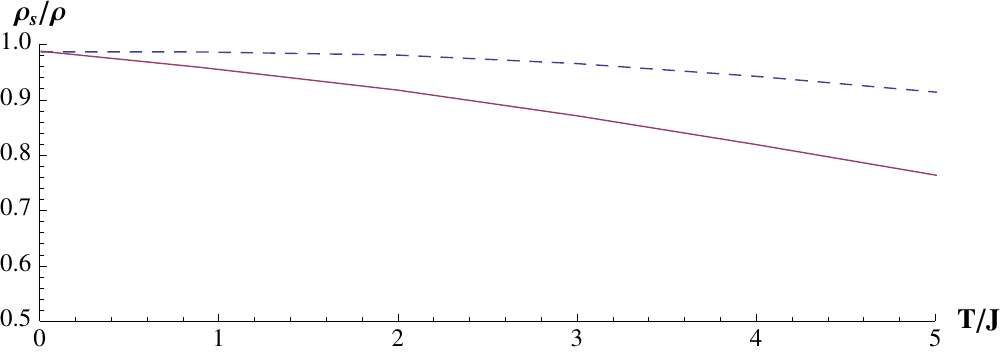} 
   }
   \subfigure[$U/J = 10$]{
   \includegraphics[width=\columnwidth]{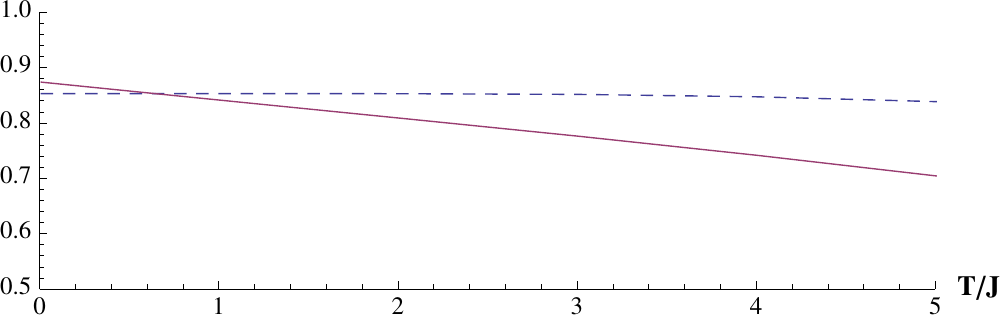} 
   }
   \caption{(Color online) The superfluid fraction $\rho_{s}/\rho$ as a function of temperature, for a two-dimensional two-by-two lattice with 16 particles, at two values of $U/J$. The dashed blue line shows the numerically exact result and the solid red line shows the analytic approximation.}
   \label{fig:2DT}
\end{figure}

We find that at zero temperature the approximate analytic expressions for the superfluid density match the numerical result well even at a relatively small number of particles per site, $\avg{n} = 4$. Moreover the agreement persists to relatively large $U$. One such example is shown in Fig.~\ref{fig:2DT0}. The finite temperature values do not agree as well with the numerical result, except for very large values of $\avg{n}$, but they follow the same trend as the numerically calculated results (see Fig.~\ref{fig:2DT}). Overall the numerical results confirm the limits of validity in Eq.~(\ref{eq:valid}).

For these comparisones we replaced the integrals in Eq.~(\ref{eq:n0})-(\ref{eq:nU}) with sums, corresponding to the finite size system.

\section{Two-Component Systems}\label{sec:2s}

\subsection{Model}

We apply the same path integral method to a system of two species of bosons on a lattice. The Hamiltonian for this system is given by
\begin{equation} \begin{split}
\hat H = \hat H_{\up} + \hat H_{\dn} + \hat H_{\up\dn}
\end{split} \end{equation}
where $\hat H_{\gs}$ for $\gs = \up,\dn$ are the single-particle Hamiltonians for particle species $\up,\dn$ respectively, identical to Eq.~(\ref{eq:BH}) except with constants $J_{\gs}, U_{\gs}, \mu_{\gs}$ and operators $\hat a_{i}^{\gs}, \p{\hat a_{i}^{\gs}}\dg, \hat n_{i}^{\gs}$ as appropriate. The final term
\begin{equation} \begin{split}
\hat H_{\up\dn} & = \sum_{i} U_{\up\dn}\hat n^{\up}_{i} \hat n^{\dn}_{i}
\end{split} \end{equation}
is the inter-species interaction Hamiltonian.

The action for this Hamiltonian given by
\begin{equation} \begin{split}
S_{E} = &\; \sum_{n} \p{\frac{a_{0}}{2\pi}}^{D}  \intrm{d^{D}k}
\\ & \br{\sum_{\gs}\p{\frac{V}{{a_{0}}^{D}}{}_{\gs}\eL_{0} + {}_{\gs}\eL_{2}^{\vk,\omega_{n}} + {}_{\gs}\eL_{int}^{\vk,\omega_{n}}}}
\\ & \quad + {}_{\up\dn}\eL_{2}^{\vk,\omega_{n}} + {}_{\up\dn}\eL_{int}^{\vk,\omega_{n}}
\end{split} \end{equation}
where ${}_{\gs}\eL_{0,2,int}$ are again identical to those defined in Eq.~(\ref{eq:L0}), (\ref{eq:Lint}), (\ref{eq:Lk}) with mean densities and phase twists $\bar\rho_{\gs}$, $\Delta\Phi^{\gs}_{d}$ substituted for the one-component equivalents as appropriate, and the saddlepoint relation
\begin{equation} \begin{split}
\mu_{\gs} & = U_{\gs}\bar\rho_{\gs} + U_{\up\dn}\bar\rho_{\bar\gs} - 2J_{\gs}\sum_{d}\cos\p{\Delta\Phi^{\gs}_{d}}
\end{split} \end{equation}
used, with $\bar\gs$ indicating the non-$\gs$ species, so that $\bar\up = \dn, \bar\dn = \up$.  The additional terms are
\begin{widetext}
\begin{equation} \begin{split}
{}_{\up\dn}\eL_{2}^{\vk,\omega}  =  2\times\half U_{\up\dn}\Delta t & \bmat{ \p{\frac{1 + \cos\p{\omega\Delta t}}{2}}\gd\rho^{\up}_{\vk,\omega} \gd\rho^{\dn}_{-\vk,-\omega}
 - 2 \bar\rho_{\up}\bar\rho_{\dn}\p{1 - \cos\p{\omega \Delta t}}\phi^{\up}_{\vk,\omega}\phi^{\dn}_{\vk,\omega} 
\\ + \sin\p{\omega \Delta t}\bar\rho_{\dn}\gd\rho^{\up}_{\vk,\omega}\phi^{\dn}_{-\vk,\omega} 
 + \sin\p{\omega \Delta t}\bar\rho_{\up}\gd\rho^{\dn}_{\vk,\omega}\phi^{\up}_{-\vk,\omega} },
\end{split} \end{equation}
and the higher order terms scale as ${}_{\up\dn}\eL_{int}^{\omega_{n},\vk} = \bar\rho_{\up}\bar\rho_{\dn}U_{\up\dn}\times O\p{1/\sqrt{\bar\rho}}^{3}$.

The in-species propagators are now, at $\omega\Delta t\ll 1$,
\begin{equation} \begin{split}
\avg{\gd\rho^{\gs}\gd\rho^{\gs}}^{p}_{\vec k,\omega} & = \frac{V}{a_{0}^{D}} \bar\rho_{\gs}
\br{\frac{1}{\Delta t}\frac{2 \E_{\gs1k}\p{\omega^{2} + {\E_{\bar\gs k}}^{2}}}{\p{\omega^{2} + {\E_{+k}}^{2}}\p{\omega^{2} + {\E_{-k}}^{2}}} + O\p{\Delta t}^{0}}
\\ \avg{\gd\rho^{\gs}\phi^{\gs}}^{p}_{\vec k,\omega}  & = \frac{V}{a_{0}^{D}}
\br{-\frac{1}{\Delta t}\frac{\omega\p{\omega^{2} + {\E_{\bar\gs k}}^{2}}}{\p{\omega^{2} + {\E_{+k}}^{2}}\p{\omega^{2} + {\E_{-k}}^{2}}} + O\p{\Delta t}^{0}}
 \\ \avg{\phi^{\gs}\phi^{\gs}}^{p}_{\vec k,\omega}  &= \frac{V}{a_{0}^{D}} \frac{1}{4\bar\rho}
 \br{\frac{1}{\Delta t}\frac{2 \E_{2k}\p{\omega^{2} + {\E_{\bar\gs k}}^{2}} - 8\bar\rho_{\up}\bar\rho_{\dn}{U_{\up\dn}}^{2}\E_{\bar\gs1k}}
 {\p{\omega^{2} + {\E_{+k}}^{2}}\p{\omega^{2} + {\E_{-k}}^{2}}} + O\p{\Delta t}^{0}}%
\label{eq:prop2spole}
\end{split} \end{equation}
while the contour pieces are identical to the single-component case. $\E_{\gs k},\E_{\gs 1k},\E_{\gs 1k}$ are the single-particle dispersion relations given in Eq.~(\ref{eq:Ek}), and the new dispersion relations are given by
\begin{equation} \begin{split}
{\E_{\pm k}}^{2} = \frac{{\E_{\up k}}^{2} + {\E_{\dn k}}^{2}}{2}  \pm \sqrt{\p{\frac{{\E_{\up k}}^{2} - {\E_{\dn k}}^{2}}{2}}^{2} 
+ 4 \bar\rho_{\up}\bar\rho_{\dn}{U_{\up\dn}}^{2} \E_{\up 1k}\E_{\dn 1k}}.
\end{split} \end{equation}
The interspecies propagators are given by
\begin{equation} \begin{split}
\avg{\gd\rho^{\up}\gd\rho^{\dn}}^{p} _{\vec k,\omega}  = \frac{V}{a_{0}^{D}}U_{\up\dn}
\br{ -\frac{1}{\Delta t}\frac{4 \bar\rho_{\up}\bar\rho_{\dn}\E_{1\up}\E_{1\dn}}{\p{\omega^{2}+{\E_{+}}^{2}}\p{\omega^{2}+{\E_{-}}^{2}}}  + O\p{\Delta t}^{0}}
&\quad\quad \avg{\gd\rho^{\up}\gd\rho^{\dn}}^{\circ}_{\vec k,\omega}  = \frac{V}{a_{0}^{D}}U_{\up\dn}\br{O\p{\Delta t}^{2}}
\\ \avg{\gd\rho^{\gs}\phi^{\bar\gs}}^{p}_{\vec k,\omega} = \frac{V}{a_{0}^{D}} U_{\up\dn}
\br{\frac{1}{\Delta t}\frac{2\bar\rho_{\gs}\E_{1\gs} \omega}{\p{\omega^{2}+{\E_{+}}^{2}}\p{\omega^{2}+{\E_{-}}^{2}}} + O\p{\Delta t}^{0}}
&\quad\quad \avg{\gd\rho^{\gs}\phi^{\bar\gs}}^{\circ}_{\vec k,\omega}  = \frac{V}{a_{0}^{D}}U_{\up\dn}\br{O\p{\Delta t}^{2}}
\\ \avg{\phi^{\up}\phi^{\dn}}^{p}_{\vec k,\omega} = \frac{V}{a_{0}^{D}} U_{\up\dn}
\br{\frac{1}{\Delta t}\frac{\omega^{2}}{\p{\omega^{2}+{\E_{+}}^{2}}\p{\omega^{2}+{\E_{-}}^{2}}} + O\p{\Delta t}^{0}}
\\  \avg{\phi^{\up}\phi^{\dn}}^{\circ}_{\vec k,\omega}  & = \frac{V}{a_{0}^{D}}U_{\up\dn}
\br{\half \frac{\sin^{2}\p{\frac{\pi}{2} e^{i\chi}}\Delta t}{\p{1 - \cos\p{\pi e^{i\chi}}}^{2}} + O\p{\Delta t}^{2}}.
\end{split} \end{equation}
\end{widetext}

\subsection{Superfluid Density}

In the presence of two species there are now three superfluid densities,
\begin{equation} \begin{split}
\rho_{s}^{\gs\tau}
= \frac{m_{\gs}m_{\tau}a_{0}^{2}}{\hbar^{2}}\br{\dee{^{2}\mathcal F}{\Delta\Phi^{\gs}_{d}\partial\Delta\Phi^{\tau}_{d}}}_{\vec{\Delta\Phi^{\gs}}=\vec{\Delta\Phi^{\tau}}=0},
\end{split} \end{equation}
where $\rho_{s}^{\gs\tau}$ is the superfluid response of species $\gs$ to the twisting of the phase of species $\tau$. The diagonal terms $\rho_{s}^{\gs\gs}$ are the superfluid densities of species $\gs$, while the off-diagonal term $\rho_{s}^{\up\dn} = \rho_{s}^{\dn\up}$ is the cross-stiffness.

The full expressions for all three terms may be calculated in a similar manner to the single-species case, as described in Appendix~\ref{app:Calcs}. At zero temperature, the superfluid densities are given by
\begin{equation} \begin{split}
\rho_{s}^{\gs} = \frac{2m_{\gs}a_{0}^{2}J_{\gs}}{\hbar^{2}} \frac{m_{\gs}}{a_{0}^{D}}\br{\avg{n^{\gs}} - n_{n}^{\gs U}}
\end{split} \end{equation}
where the number of normal atoms per site is given by
\begin{equation} \begin{split}
& n^{\gs U}_{n}  = \half  \intrm{\frac{a_{0}^{D}d^{D}k}{\p{2\pi}^{D}}} \p{1 - \cos\p{k_{j}a_{0}}} \times
\\ & \quad \bmat{
 \frac{\p{\E_{\gs1k}+\E_{\gs2k}}\p{\E_{+k}\E_{-k} + {\E_{\bar\gs k}}^{2}} - 4 \bar\rho_{\up}\bar\rho_{\dn} {U_{\up\dn}}^{2}\E_{\gs 1k}}{2\E_{+k}\E_{-k}\p{\E_{+k} + \E_{-k}}} - 1}.
\end{split} \end{equation}
The cross-stiffness at $T = 0$ is
\begin{equation} \begin{split}
\rho_{s}^{\up\dn} &= \frac{2\sqrt{m_{\up}m_{\dn}}a_{0}^{2}\sqrt{J_{\up} J_{\dn}}}{\hbar^{2}} \frac{\sqrt{m_{\up}m_{\dn}}}{a_{0}^{D}} \times
\\ &\intrm{\frac{a_{0}^{D}d^{3}k}{\p{2\pi}^{3}}}  \sin^{2}\p{k_{d}a_{0}}\frac{4 \sqrt{J_{\up}J_{\dn}} \bar\rho_{\up}\bar\rho_{\dn}U_{\up\dn}^{2} \E_{1\up}\E_{1\dn}}
 {\E_{+} \E_{-}\p{\E_{+} + \E_{-}}^{3}}.
\end{split} \end{equation}

\begin{figure}[htbp] 
   \centering
   \includegraphics[width=\columnwidth]{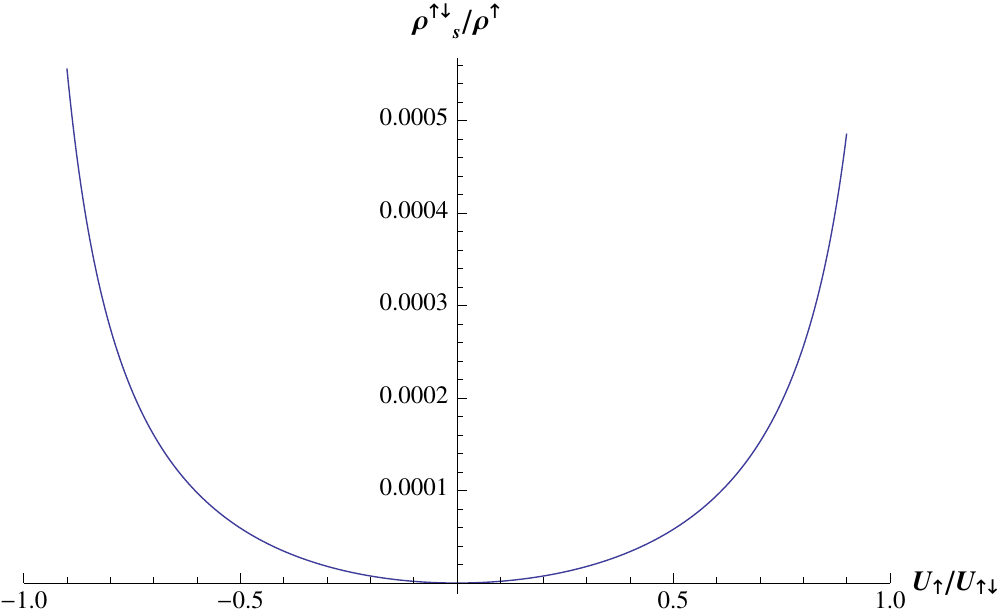} 
   \caption{The superfluid cross-stiffness $\rho_{s}^{\up\dn}/\rho^{\up}$ as function of the interspecies interaction $U_{\up\dn}/U_{\up}$ for a two-component Bose gas on an infinite 3D cubic lattice. Here $\avg{n^{\up}} = \avg{n^{\dn}} = 10$, with $U_{\dn} = U_{\up} = 10 J_{\dn} = 10 J_{\up}$, calculated to leading order in a $1/\avg{n}$ expansion.}
   \label{fig:rho12}
\end{figure}

While the cross-stiffness is expected to be substantial in hard-core bosons \cite{Kaurov2005} it is negligible in the weak-interaction case, as illustrated in Fig.~\ref{fig:rho12}.

\begin{figure}[htbp] 
   \centering
   \includegraphics[width=\columnwidth]{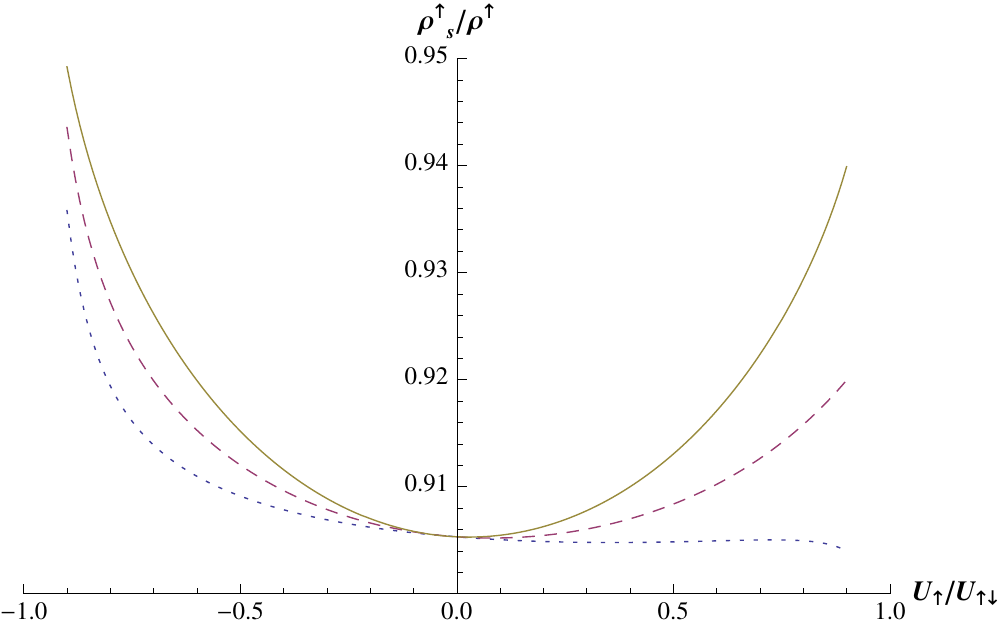} 
   \caption{(Color online) The superfluid fraction of the one component $\rho^{\up}_{s}/\rho^{\up}$ as function of the interspecies interaction $U_{\up\dn}/U_{\up}$ for a two-component Bose gas on an infinite 3D cubic lattice. Here $\avg{n^{\up}} = \avg{n^{\dn}} = 10$,  $U_{\dn}=U_{\up} = 10J_{\up}$, and the second component has hopping parameter $J_{\dn}/J_{\up} = 0.1$  (dotted blue line), $J_{\dn}/J_{\up} = 1$ (dashed red line) and $J_{\dn}/J_{\up}=10$ (solid yellow line).}
   \label{fig:3Drho1U12}
\end{figure}

A more dramatic effect can be seen in the superfluid densities. At zero temperature, a strong coupling to a second species of particles can replenish the superfluid fraction, as long as the superfluid has an equal or larger hopping parameter. This is seen in Fig.~\ref{fig:3Drho1U12}.

\section{Outlook}

The $T = 0$ normal density, $\rho_{n}$, for lattice bosons is generally non-zero. As discussed in Sec. \ref{sec:singlesp}, this property, and the temperature dependence of the superfluid density, can be experimentally studied using cold atoms. Here we calculated $\rho_{n}$, and proposed comparing our results with experiment.

For our calculation we extended the standard saddle-point functional integral approach. When using a coherent state basis, the discrete time path integral contains extra terms over the continuous time limit version. We explicitly derived those corrections for the Bose Hubbard model. Similar issues appear in spin models, and our techniques could be applied there.

Our results are applicable at high density, low temperature, and weak interaction. One could envision extending them to strong interaction by using a different set of coherent states. For example, in the hard core limit it would be natural to use $\ket{\theta,\varphi}_{i} = \cos\theta \ket{0}_{i} + e^{i\varphi}\sin\theta \ket{1}_{i}$, where $\ket{0}_{i}, \ket{1}_{i}$ are the states with no particles or one particle on site $i$ respectively. The other approach to extending the validity of our results would be to include perturbative corrections. In particular, one might envision summing an infinite set of these corrections using Feynman diagram techniques.

We also present results for the superfluid properties of two-component lattice bosons. These are an active area of research, and there are rich possibilities for exploring our formalism in those systems. One experiment \cite{Gadway2010} has seen hints of the impact of one bosonic species on the superfluid properties of another. Those results appear to be in the opposite direction from our predictions - however they are in a stronger interacting regime, near the superfluid-to-Mott insulator transition and the quantitative applicability of our results to their experiment is questionable. We also neglect any processes which involve higher bands.

\section*{Acknowledgements}

This paper is based upon work supported by the National Science Foundation under Grant No. PHY-1068165 and by a grant from the Army Research Office with funding from the DARPA OLE program.

\appendix
\section{Discrete Time Path Integrals}\label{app:DiscTPT}

The traditional path integral formulation of quantum mechanics involves the transformation of the partition function
\begin{equation} \begin{split}
Z & = \Tr e^{-\gb \hat H} = \sum_{\ket{\psi}}\bra{\psi} e^{-\gb \hat H} \ket{\psi}
\end{split} \end{equation}
into a path integral. Here $\acom{\ket{\psi}}$ is any complete basis of the states. In our case we will use the overcomplete basis of coherent states.

To make the transformation, we break up the operator $e^{-\gb \hat H} = \br{e^{-\gb \hat H/N_{t}}}^{N_{t}}$ into $N_{t}$ time steps. We then insert an identity operator $\mathbb{\hat 1} = \sum_{\ket{\psi}}\ket{\psi}\bra{\psi}$ between each step,
\begin{equation} \begin{split}
Z = \sum_{\acom{\ket{\psi_{t}}}} \prod_{t} \bra{\psi_{t}} e^{-\gb \hat H/N_{t}} \ket{\psi_{t+1}}
\end{split} \end{equation}
where the summation is over $N_{t}$ copies of the the basis $\ket{\psi}$, the product is over $t = 0..N_{t}-1$ and we define $\ket{\psi_{N_{t}}} = \ket{\psi_{0}}$.

Taking the number of time steps $N_{t} \ll 1$ to be very large, we then expand
\begin{equation} \begin{split}
& \bra{\psi_{t}} e^{-\gb \hat H/N_{t}} \ket{\psi_{t+1}} =
\\ &\quad \braket{\psi_{t}}{\psi_{t+1}} - \frac{\gb}{N_{t}}\bra{\psi_{t}} \hat H \ket{\psi_{t+1}} + O\p{\gb/N_{t}}^{2}
\\ & \quad \approx \exp\br{\log\br{\braket{\psi_{t}}{\psi_{t+1}}} - \frac{\gb}{N_{t}}\frac{\bra{\psi_{t}} \hat H \ket{\psi_{t+1}}}{\braket{\psi_{t}}{\psi_{t+1}}} }.
\end{split} \end{equation}
The integration over all $N_{t}$ values of $\ket{\psi_{t}}$ is a path integral, and one finds
\begin{equation} \begin{split}
Z = \intrm{\D \psi}e^{-S_{E}}
\end{split} \end{equation}
where $S_{E} =  \sum_{t} L_{t}$ and
\begin{equation} \begin{split}
L_{t}  = -\log\br{\braket{\psi_{t}}{\psi_{t+1}}}& + \frac{\gb}{N_{t}}\frac{\bra{\psi_{t}} \hat H \ket{\psi_{t+1}}}{\braket{\psi_{t}}{\psi_{t+1}}}.
\label{eq:LEvalid}
\end{split} \end{equation}

It is here that we diverge from the tradition continuous formulation of the path integral. One typically assumes
\begin{equation} \begin{split}
\ket{\psi_{t+1}} = \p{1 + \p{\gb/N_{t}} \partial_{t} + O\p{\gb/N_{t}}^{2}}\ket{\psi_{t}},
\label{eq:contapprox}
\end{split} \end{equation}
and by taking $N_{t}\to\infty$ the sum can then be converted into an integral, $S_{E}^{cont} = \intrm{dt} L_{t}$, where
\begin{equation} \begin{split}
L_{t} = - \bra{\psi\p{t}}\partial_{t} \ket{\psi\p{t}} + \bra{\psi\p{t}}\hat H \ket{\psi\p{t}}.
\end{split} \end{equation}
The expression in Eq.~(\ref{eq:contapprox}) is not always valid. In particular, for an overcomplete basis the overlap $\braket{\psi_{t}}{\psi_{t+1}}$ remains finite for states that differ to a non-infinitesimal degree, and the difference between $\ket{\psi_{t+1}}$ and $\ket{\psi_{t}}$ need not go to zero as $N_{t}\to \infty$. This leads to a breakdown of the traditional continuous path integral, as shown in \cite{Wilson2011}. However, the discrete time formulation Eq.~(\ref{eq:LEvalid}) remains valid.

One example is the single-site Bose-Hubbard model,
\begin{equation} \begin{split}
\hat H_{ss} = \frac{U}{2}\hat n \p{\hat n-1} - \mu \hat n
\end{split} \end{equation}
where $\hat n$ is the number operator.

The partition function for this Hamiltonian can be calculated in the Fock basis,
\begin{equation} \begin{split}
Z_{ss} = \sum_{n}\exp\br{-\gb\p{\frac{U}{2} n\p{n-1} - \mu n}}.
\end{split} \end{equation}
At $T = 0$, the mean occupation number is then the integer $n$ that minimizes the exponent,
\begin{equation} \begin{split}
\avg{n} = \frac{\mu}{U} + \half.
\label{eq:nFock}
\end{split} \end{equation}

Following the continuous time path integral formalism yields the wrong result for the partition function,
\begin{equation} \begin{split}
Z_{ss}\pr = \sum_{n}\exp\br{-\gb\p{\frac{ U}{2} n^{2} + \mu n}},
\end{split} \end{equation}
and hence the wrong result of $\avg{n} = \mu/U$.

However, application of the discrete time path integral Eq.~(\ref{eq:LEvalid}) yields the correct value.  Using a coherent state basis one finds
\begin{equation} \begin{split}
L_{t} & =  -\log\br{\braket{\rho_{t}\varphi_{t}}{\rho_{t+1}\varphi_{t+1}}} 
\\ & \quad + \Delta t\frac{\bra{\rho_{t}\varphi_{t}}\hat H_{ss}\ket{\rho_{t+1}\varphi_{t+1}}}{\braket{\rho_{t}\varphi_{t}}{\rho_{t+1}\varphi_{t+1}}}
\\ & = \half\p{\rho_{t} + \rho_{t+1}} - \sqrt{\rho_{t}\rho_{t+}}e^{i\p{\varphi_{t+1} - \varphi_{t}}}
\\ &\quad + \frac{U \Delta t}{2}\rho_{t}\rho_{t+1}e^{2i\p{\varphi_{t+1} - \varphi}} 
\\ & \quad - \mu \Delta t \sqrt{\rho_{t}\rho_{t+1}}e^{i\p{\varphi_{t+1} - \varphi}}.
\end{split} \end{equation}
where $\Delta t = \gb/N_{t}$. Using a saddleploint approximation,
\begin{equation} \begin{split}
\rho_{t} = \bar\rho + \gd\rho_{t},
\end{split} \end{equation}
we have
\begin{equation} \begin{split}
Z = &\intrm{\D\gd\rho\D\phi}
\\ &\quad \exp\br{-\sum_{t}\eL_{0} + \eL_{1}^{t} + \eL_{2}^{t} + \eL_{int}^{t}},
\end{split} \end{equation}
where
\begin{equation} \begin{split}
\eL_{0} = \p{\frac{U}{2}\bar\rho^{2} - \mu \bar\rho}\Delta t,
\end{split} \end{equation}
\begin{equation} \begin{split}
\eL_{1} & = \p{U \bar \rho - \mu}\Delta t\gd\rho_{t},
\end{split} \end{equation}
\begin{equation} \begin{split}
\eL_{2} & = \half \br{1 - \p{2U \bar\rho - \mu}\Delta t} \times
\\ &\quad\quad \bmat{ \frac{\p{\gd\rho_{\tau+1} - \gd\rho_{\tau}}^{2}}{4\bar\rho} + \bar\rho\p{\varphi_{\tau+1} - \varphi_{\tau}}^{2}
\\ -i \p{\gd\rho_{\tau+1} + \gd\rho_{\tau}}\p{\varphi_{\tau+1} - \varphi_{\tau}} }
\\ & \quad + \frac{1}{4} U\Delta t \p{\gd\rho_{t}^{2} + \gd\rho_{t+1}^{2}}.
\end{split} \end{equation}

We choose $\bar\rho = \mu/U$, as in the continuous time integral case, to minimize the zeroth-order action. Then in momentum space
\begin{equation} \begin{split}
Z = \intrm{\D\gd\rho\D\phi}\exp\br{-\sum_{\omega}\eL_{0} + \eL_{2}^{\omega} + \eL_{int}^{\omega}}
\label{eq:ssZw}
\end{split} \end{equation}
where the sum is over $\omega = -\frac{2\pi}{\gb}\frac{N_{t}-1}{2}\dotsc\frac{2\pi}{\gb}\frac{N_{t}-1}{2}$. Here, after the substitution, $\eL_{0} = -\half \frac{\mu^{2}\Delta t}{U}$ and
\begin{equation} \begin{split}
\eL_{2} & = \half \mat{ \gd\rho_{\omega} & \varphi_{\omega} } G^{-1}_{\omega}\mat{ \gd\rho_{-\omega} \\ \varphi_{-\omega} }
\label{eq:ssL2}
\end{split} \end{equation}
where
\begin{equation} \begin{split}
\br{G^{-1}_{\omega}}_{1,1} & = \frac{ \p{1 - \cos\p{\omega\Delta t}} + \p{1 + \cos\p{\omega\Delta t}}\mu \Delta t}{2\mu/U}
\\ \br{G^{-1}_{\omega}}_{1,2} & = -\br{G^{-1}_{\omega}}_{2,1} = -\sin\p{\omega\Delta t}\p{1 - \mu\Delta t}
\\ \br{G^{-1}_{\omega}}_{2,2} & = -2\p{\mu/U}\p{1-\cos\p{\omega\Delta t}}\p{1 - \mu\Delta t}.
\end{split} \end{equation}
We invert $G^{-1}$ to find the propagators
\begin{equation} \begin{split}
\avg{\gd\rho_{\omega}\gd\rho_{\eta}} & = \gd_{\omega,-\eta}\mu/U
\\ \avg{\gd\rho_{\omega}\varphi_{\eta}} & = -\gd_{\omega,-\eta}\half \cot\p{\omega\Delta t/2}
\\ \avg{\varphi_{\omega}\varphi_{\eta}} & = 
\\ \gd_{\omega,-\eta}\frac{U}{4\mu}&\br{\frac{\p{1 + \mu \Delta t} - \cos\p{\omega\Delta t}\p{1 - \mu \Delta t}}{\p{1 - \mu \Delta t}\p{1 - \cos\p{\omega\Delta t}}}}.
\end{split} \end{equation}

Next we calculate
\begin{equation} \begin{split}
\avg{n} & = -\dee{F}{\mu} = - \frac{1}{\gb}\sum_{\omega}\avg{\dee{\eL_{0}}{\mu}} + \avg{\dee{\eL_{2}^{\omega}}{\mu}},
\label{eq:ssprop}
\end{split} \end{equation}
where we have neglected $\eL_{int}^{\omega}$. Inserting the values of the propagators in Eq.~(\ref{eq:ssprop}) into the derivative of Eqs.~(\ref{eq:ssZw})-(\ref{eq:ssL2}), we find
\begin{equation} \begin{split}
\avg{n} 
& = 
\\ \frac{\mu}{U} + & \frac{1}{\gb}\sum_{\omega}\bmat{\frac{1 - \cos\p{\omega\Delta t}}{4\mu}
 + \half\Delta t\sin\p{\omega \Delta t} \cot\p{\omega\Delta t/2}
\\ - \frac{\p{1 - 2\mu\Delta t}}{4\mu}\br{\frac{\p{1 + \mu \Delta t} - \cos\p{\omega\Delta t}\p{1 - \mu \Delta t}}{\p{1 - \mu \Delta t}}}}
\\ & = \frac{\mu}{U} + \frac{1}{\gb}\sum_{\omega}\half \frac{\Delta t}{1 - \mu \Delta t} = \frac{\mu}{U} + \half \frac{1}{1 - \mu \Delta t}.
\end{split} \end{equation}

As we take $\Delta t \to 0$ this becomes $\avg{n} = \mu/U + \half$, in agreement with Eq.~(\ref{eq:nFock}). A subtle error still remains, namely that at zero temperature the occupation number must be an integer. To restore this constraint one would need to explicit sum over the topologically distinct ``instanton'' paths where $\varphi$ has multiple windings.

\section{Explicit Calculations in the Discrete Time Path Integral}\label{app:Calcs}

We provide here a further explicit example of a calculation in the discrete time step path integral formalism. For a more elementary example see Appendix \ref{app:DiscTPT}. The most important result in this appendix is the development of a formalism in which a discrete time calculation is expressed as the sum of a continuous time one and some easily calculated corrections.

Our starting point is Eq.~(\ref{eq:SEk}). Explicitly, the quadratic term is
\begin{equation} \begin{split}
\eL_{2} & = \half \mat{ \frac{\gd\rho_{\vk,\omega}}{2\sqrt{\bar\rho}} & \sqrt{\bar\rho}\phi_{\vk,\omega} } 
\mathcal G^{-1}_{\vk,\omega}\mat{ \frac{\gd\rho_{-\vk,-\omega}}{2\sqrt{\bar\rho}} \\ \sqrt{\bar\rho}\phi_{-\vk,-\omega} },
 \label{eq:Lk}
\end{split} \end{equation}
where $\mathcal G^{-1}$ is an inverse Green's function matrix,
\begin{widetext}
\begin{equation} \begin{split}
& \mathcal G^{-1}_{\vk,\omega} = \\
&\mat{
\begin{array}{c}
2\p{1 - \cos\p{\omega \Delta t}} + 2\p{1 + \cos\p{\omega \Delta t}}\bar\rho U \Delta t
 \\ + 4J \Delta t \cos\p{\omega \Delta t}\sum_{d}\p{1 - \cos\p{k_{d}a_{0}}}\cos\p{\Delta\Phi_{d}}
 \\ + 4 i J\Delta t \sin\p{\omega\Delta t} \sum_{d} \sin\p{k_{d} a_{0}} \sin\p{\Delta\Phi_{d}}
\end{array}
 &
\begin{array}{c}
-2\sin\p{\omega \Delta t}\p{1 - \bar\rho U \Delta t}
 \\ + 4 J \Delta t \sin\p{\omega \Delta t}\sum_{d}\p{1 - \cos\p{k_{d}a_{0}}}\cos\p{\Delta\Phi_{d}}
 \\ - 4 i J\Delta t \cos\p{\omega\Delta t} \sum_{d} \sin\p{k_{d} a_{0}} \sin\p{\Delta\Phi_{d}}
\end{array}
\\
\\ 
\begin{array}{c}
2\sin\p{\omega \Delta t}\p{1 - \bar\rho U \Delta t}
 \\ - 4 J \Delta t \sin\p{\omega \Delta t}\sum_{d}\p{1 - \cos\p{k_{d}a_{0}}}\cos\p{\Delta\Phi_{d}}
 \\ + 4 i J\Delta t \cos\p{\omega\Delta t} \sum_{d} \sin\p{k_{d} a_{0}} \sin\p{\Delta\Phi_{d}}
\end{array}
&
\begin{array}{c}
2\p{1 - \cos\p{\omega \Delta t}}\p{1 - \bar\rho U\Delta t}
 \\ + 4J\Delta t \cos\p{\omega\Delta t} \sum_{d}\p{1 - \cos\p{k_{d}a_{0}}} \cos\p{\Delta\Phi_{d}}
 \\ + 4 i J\Delta t \sin\p{\omega\Delta t} \sum_{d} \sin\p{k_{d} a_{0}} \sin\p{\Delta\Phi_{d}}
\end{array}
 }.
\end{split} \end{equation}

The propagators are obtained by inverting $\mathcal G^{-1}$, and are given by
\begin{equation} \begin{split}
\avg{\gd\rho_{\vec k,\omega} \gd\rho_{\vec q,\eta}} & = \frac{\gd^{\p{D}}\p{\vec q + \vec k}\gd_{\omega,-\eta}}{\p{a_{0}/2\pi}^{D}} \br{
\bar\rho\frac{2\p{1 - \cos\p{\omega\Delta t}} + \br{\p{\E_{1k} + \E_{2k}}\cos\p{\omega\Delta t} + \p{\E_{1k} - \E_{2k}}}\Delta t}
{2\p{1 - \cos\p{\omega \Delta t}}\p{1 - \half\p{\E_{1k} + \E_{2k}}\Delta t} + {\E_{k}}^{2}\Delta t^{2}} + O\p{\Delta\Phi}}
\\ \avg{\gd\rho_{\vec k,\omega} \phi_{\vec q,\eta}} & = \frac{\gd^{\p{D}}\p{\vec q + \vec k}\gd_{\omega,-\eta}}{\p{a_{0}/2\pi}^{D}} \br{
-\sin\p{\omega\Delta t}\frac{1 - \half \p{\E_{1k} +  \E_{2k}}\Delta t}{2\p{1 - \cos\p{\omega \Delta t}}\p{1 - \half\p{\E_{1k} + \E_{2k}}\Delta t} + {\E_{k}}^{2}\Delta t^{2}}
 + O\p{\Delta\Phi}}
\\ \avg{\phi_{\vec k,\omega} \phi_{\vec q,\eta}} & = \frac{\gd^{\p{D}}\p{\vec q + \vec k}\gd_{\omega,-\eta}}{\p{a_{0}/2\pi}^{D}} \br{
\frac{1}{4\bar\rho}\frac{2\p{1 - \cos\p{\omega\Delta t}} + \br{\p{\E_{1k} + \E_{2k}}\cos\p{\omega\Delta t} + \p{\E_{2k} - \E_{1k}}}\Delta t}
{2\p{1 - \cos\p{\omega \Delta t}}\p{1 - \half\p{\E_{1k} + \E_{2k}}\Delta t} + {\E_{k}}^{2}\Delta t^{2}} + O\p{\Delta\Phi}}.
\label{eq:prop}
\end{split} \end{equation}
\end{widetext}

As explained below, the only important features of these functions are their  $\omega\Delta t \to 0$ structures and their values at $\abs{\omega\Delta t} = \pi$. We illustrate this result by calculating the average occupation number via
\begin{equation} \begin{split}
\frac{\avg{n}}{a_{0}^{D}} & = - \dee{\mathcal F}{\mu} 
= \frac{1}{\gb V} \frac{1}{Z} \dee{Z}{\mu} = 
\\ & -\frac{1}{\gb V}\p{\frac{a_{0}}{2\pi}}^{D} \intrm{d^{D}k} \sum_{\omega_{n}} 
\\ & \; \br{\frac{V}{{a_{0}}^{D}}\avg{\dee{\eL_{0}}{\mu}} + \avg{\dee{\eL_{2}^{\vk,\omega_{n}}}{\mu}} + \avg{\dee{\eL_{int}^{\vk,\omega_{n}}}{\mu}}}.
\label{eq:appBn}
\end{split} \end{equation}
As we have assigned $\bar\rho = \frac{1}{U}\p{\mu + 2J\sum_{d}\cos\p{\Delta\Phi_{d}}}$, the derivatives are given by
\begin{equation} \begin{split}
\dee{}{\mu} = \at{\dee{}{\mu}}{\bar\rho} + \dee{\bar\rho}{\mu}\at{\dee{}{\bar\rho}}{\mu}.
\end{split} \end{equation}
We calculate this quantity at $\vec{\Delta\Phi} = 0$. 

The saddle point contribution comes from the constant
\begin{equation} \begin{split}
\dee{\eL_{0}}{\mu} & = -\bar\rho \Delta t.
\end{split} \end{equation}
The contribution from this term to Eq.~(\ref{eq:appBn}) is
\begin{equation} \begin{split}
-\frac{1}{\gb V}\p{\frac{a_{0}}{2\pi}}^{D} &\intrm{d^{D}k} \sum_{\omega_{n}} \frac{V}{{a_{0}}^{D}}\avg{\dee{\eL_{0}}{\mu}} = \frac{\bar\rho}{a_{0}^{D}}.
\label{eq:appBnrho}
\end{split} \end{equation}

The nontrivial part of the calculation comes from the term involving $\eL_{2}^{\vk,\omega}$,
\begin{equation} \begin{split}
\dee{\eL_{2}^{\vk,\omega_{n}}}{\mu} 
& = \frac{1}{\mu}\p{1 - \cos\p{\omega\Delta t} + \E_{1k} \Delta t \cos\p{\omega\Delta t}}\times
\\ &\quad\quad \p{\bar\rho\phi_{\vk,\omega}\phi_{-\vk,-\omega} - \frac{\gd\rho_{\vk,\omega}\gd\rho_{-\vk,-\omega}}{4\bar\rho} }
\\ & \quad + \sin\p{\omega\Delta t} \Delta t\gd\rho_{\vk,\omega}\phi_{-\vk,-\omega}
\\ & \quad - \p{1 - \cos\p{\omega\Delta t}} 2 \Delta t \bar\rho \phi_{\vk,\omega}\phi_{-\vk,-\omega}.
\label{eq:dL2mu}
\end{split} \end{equation}

\begin{figure}[htbp] 
   \centering
   \includegraphics[width=\columnwidth]{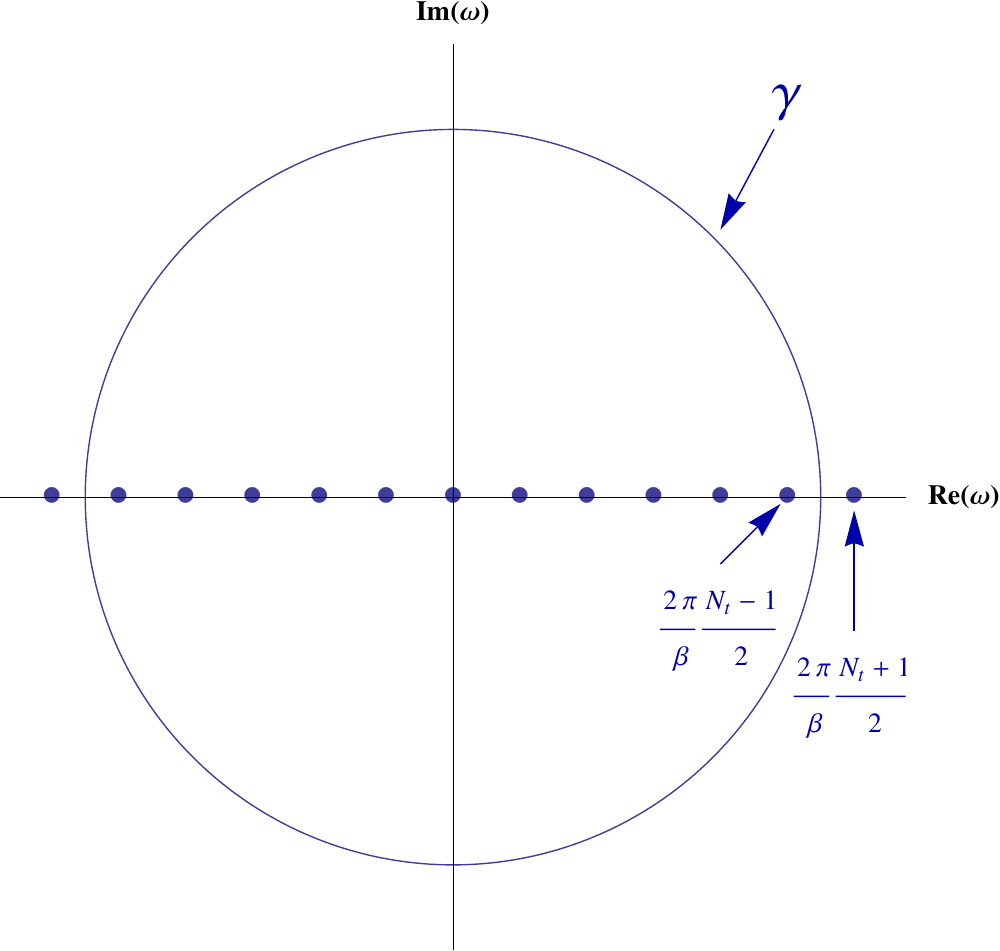} 
   \caption{The contour $\gamma$ used to perform the summation over $\omega = -\frac{2\pi}{\gb}\frac{N_{t}-1}{2}\dotsc \frac{2\pi}{\gb}\frac{N_{t}-1}{2}$ as in Eq.~(\ref{eq:sumrule}). The contour is given by $\omega = \frac{\pi}{\Delta t}e^{i\chi}$, $\chi = 0\dotsc 2\pi$. As one goes to the continuous time case with $\Delta t\to 0$, the radius of the contour goes to infinity. }
   \label{fig:contour}
\end{figure}

We perform the summation over the frequencies $\omega_{n}$ by taking a contour integral. The same trick is used in the continuous time approach, but the contour here is slightly different. As illustrated in Fig.~\ref{fig:contour}, the integration is performed over a circle of finite radius $\frac{2\pi}{\gb}\frac{N_{t}-1}{2} < \abs{\omega} <\frac{2\pi}{\gb}\frac{N_{t}+1}{2}$. In terms of the integral over this contour $\gamma$, the summation over frequencies can be expressed as
\begin{equation} \begin{split}
\frac{1}{\gb}\sum_{\omega_{n}} F\p{\omega} & = \frac{1}{2\pi} \oint_{\gamma} \mathrm{d\omega} \frac{F\p{\omega}}{e^{i\gb \omega} - 1}
\\ & \quad\quad -i\sum_{\omega_{F}}\rm{Res}\br{\frac{F\p{\omega}}{e^{i\gb \omega} - 1},\omega_{F}}.
\label{eq:sumrule}
\end{split} \end{equation}
The sum on the left hand side is over the frequencies $\omega_{n} = -\frac{2\pi}{\gb}\frac{N_{t}-1}{2}\dotsc \frac{2\pi}{\gb}\frac{N_{t}-1}{2}$, the sum on the right is over the poles $\omega_{F}$ of $F\p{\omega}$ inside the contour $\gamma$, and $\gamma$ is the complex circle defined by $\abs{\omega} = \frac{2\pi}{\gb}\frac{N_{t}}{2} = \frac{\pi}{\Delta t}$. The notation $\rm{Res}\p{X\p{\omega},\omega_{F}}$ refers to the residue of $X\p{\omega}$ at $\omega = \omega_{F}$.

In a continuous time calculation, one takes the contour $\gamma$ to infinity. Assuming $F\p{\omega}$ is well behaved, the integral on the right-hand side of Eq.~(\ref{eq:sumrule}) then vanishes. In our case we must explicitly include this term. To calculate the contour integral, we take $\omega = \frac{\pi}{\Delta t}e^{i\chi}$, with $\chi = 0\dotsc 2\pi$. As $\Delta t\to 0$, the Bose factor is
\begin{equation} \begin{split}
\p{e^{i\pi\p{\gb/\Delta t} e^{i\chi}} - 1}^{-1} \to \left\{ \begin{array}{cl}
-1 & 0 < \chi < \pi
\\ 0 & \pi < \chi < 2\pi
\end{array} \right.
\end{split} \end{equation}
and so the integral of Eq.~(\ref{eq:sumrule}) becomes
\begin{equation} \begin{split}
\oint_{\gamma} \mathrm{d\omega} \frac{F\p{\omega}}{e^{i\gb \omega} - 1}
 = -i\frac{\pi}{\Delta t}\intrml{d\chi}{0}{\pi} e^{i\chi} F\p{\frac{\pi}{\Delta t}e^{i\chi}}.
\end{split} \end{equation}

In the limit $\Delta t \to 0$, the poles of the functions in Eq.~(\ref{eq:prop}) converge to finite values of $\omega$. Hence $\omega_{F}\Delta t \ll 1$, and the sum over $\omega_{F}$ accesses only information about the low-energy structure of Eq.~(\ref{eq:prop}) and (\ref{eq:dL2mu}).

Thus, the summation of functions of the propagators in Eq.~(\ref{eq:prop}) requires only the form of their low-frequency poles, at $\omega\Delta t\ll 1$ (marked with superscript $p$) and their values on the contour $\gamma$, at $\omega = \frac{\pi}{\Delta t}e^{i\chi}$ (marked by superscript $\circ$). For our particular case, the summand Eq.~(\ref{eq:dL2mu}) is composed of 
\begin{equation} \begin{split}
\avg{\dee{\eL_{2}^{\vk,\omega_{n}}}{\mu}}^{p} & = \frac{\E_{1k}\Delta t}{\mu} 
\p{\bar\rho\avg{\phi\phi}^{p}_{\vk,\omega} - \frac{\avg{\gd\rho\gd\rho}^{p}_{\vk,\omega}}{4\bar\rho}}
\\ &  + O\p{\Delta t}^{2}\times\p{\avg{\gd\rho\gd\rho},\avg{\gd\rho\phi},\avg{\phi\phi}},
\end{split} \end{equation}
and
\begin{equation} \begin{split}
& \avg{\dee{\eL_{2}^{\vk,\omega_{n}}}{\mu}}^{\circ}   = \sin\p{\pi e^{i\chi}} \Delta t\avg{\gd\rho\phi}^{\circ}_{\vk,\omega} 
\\ & \qquad - \p{1 - \cos\p{\pi e^{i\chi}}} 2 \Delta t \bar\rho \avg{\phi\phi}^{\circ}_{\vk,\omega}
\\ & \qquad + \frac{1}{\mu} \p{1 - \cos\p{\pi e^{i\chi}} + \E_{1k} \Delta t \cos\p{\pi e^{i\chi}}}\times
\\ & \qquad\qquad \p{\bar\rho\avg{\phi\phi}^{\circ}_{\vk,\omega} - \frac{\avg{\gd\rho\gd\rho}^{\circ}_{\vk,\omega}}{4\bar\rho} }.
\end{split} \end{equation}

Thus we do not need the full structure given in Eq.~(\ref{eq:prop}) byte rather just the pole and contour values given in Eq.~(\ref{eq:proppole}). 

We now explicitly calculate the contribution of $\dee{}{\mu}\eL_{2}^{\vk,\omega_{n}}$ to $\avg{n}$. The low-frequency behavior is 
\begin{equation} \begin{split}
\avg{\dee{\eL_{2}^{\vk,\omega_{n}}}{\mu}}^{p} & = \frac{V}{a_{0}^{D}}\frac{ \E_{1k}}{\omega^{2} + \E_{k}^{2}}
\end{split} \end{equation}
with poles at $\omega_{F} = \pm i \E_{k}$, so that 
\begin{equation} \begin{split}
-i\sum_{\omega_{F}} \rm{Res}& \br{\avg{\dee{\eL_{2}^{\vk,\omega}}{\mu}}^{p}\p{e^{i\gb \omega} - 1}^{-1},\omega_{F}}
\\ & = \frac{V}{a_{0}^{D}} \half \frac{\E_{1k}}{\E_{k}}\coth\p{\gb \E_{k}/2},
\end{split} \end{equation}
 while the contour value is
\begin{equation} \begin{split}
\avg{\dee{\eL_{2}^{\vk,\omega_{n}}}{\mu}}^{\circ} & = -\frac{V}{a_{0}^{D}} \half \Delta t
\\ \frac{1}{2\pi}\oint_{\gamma} \mathrm{d\omega}\avg{\dee{\eL_{2}^{\vk,\omega_{n}}}{\mu}}^{\circ} &\p{e^{i\gb \omega} - 1}^{-1} = -\half \frac{V}{a_{0}^{D}}.
\end{split} \end{equation}
Hence
\begin{equation} \begin{split}
-\frac{1}{\gb V}&\p{\frac{a_{0}}{2\pi}}^{D} \intrm{d^{D}k} \sum_{\omega_{n}} \avg{\dee{\eL_{2}^{\vk,\omega_{n}}}{\mu}} 
\\ & = \half \intrm{\frac{d^{D}k}{\p{2\pi}^{D}}} \p{1 - \frac{\E_{1k}}{\E_{k}}\coth\p{\gb \E_{k}/2}}.
\end{split} \end{equation}

Combining this result with the zeroth-order contribution in Eq.~(\ref{eq:appBnrho}) we find
\begin{equation} \begin{split}
\\ \avg{n} & = \bar\rho + \half\intrm{\frac{a_{0}^{D}d^{D}k}{\p{2\pi}^{D}}} \p{1  - \frac{\E_{1k}}{\E_{k}}\coth\p{\gb \E_{k}/2}}
\end{split} \end{equation}
where $\bar\rho = \p{\mu + 2J D}/U$.

\bibliography{/Users/yarivyanay/Documents/University/Citations/Superfluid,/Users/yarivyanay/Documents/University/Citations/Kuklov2SF,/Users/yarivyanay/Documents/University/Citations/CoherentStatesFT}
\bibliographystyle{apsrev}

\end{document}